\documentclass[acmsmall]{acmart}

\AtBeginDocument{%
  }

\setcopyright{acmcopyright}
\copyrightyear{2023}
\acmYear{2023}
\acmDOI{XXXXXXX.XXXXXXX}

\acmJournal{TOIS}
\acmVolume{01}
\acmNumber{01}
\acmArticle{01}
\acmMonth{2}

\usepackage{natbib}
\usepackage{algorithm}
\usepackage{amsmath}
\usepackage{array}
\usepackage{booktabs}
\usepackage{subcaption}
\usepackage{graphicx}
\usepackage{multirow}
\usepackage[inline]{enumitem}
\usepackage[skip=0pt]{caption}
\usepackage{xcolor}
\usepackage{pifont}
\usepackage{acronym}

\acrodef{MII}{missing information imputation}
\acrodef{MIIR}{missing information imputation recommender}
\acrodef{DFSA}{dense fusion self-attention}
\acrodef{SFSA}{sparse fusion self-attention}
\acrodef{RS}{recommender system}
\acrodef{SR}{sequential recommendation}
\acrodef{CV}{computer vision}
\acrodef{NLP}{natural language processing}
\acrodef{SSL}{self-supervised learning}
\acrodef{RNN}{recurrent neural network}
\acrodef{GCN}{graph convolutional network}
\acrodef{AE}{auto-encoder}
\acrodef{KG}{knowledge graph}

\author{Yujie Lin}
\orcid{0000-0002-2146-0626}
\affiliation{%
\institution{Shandong University}
\city{Qingdao}
\country{China}
}
\email{yu.jie.lin@outlook.com}

\author{Zhumin Chen}
\orcid{0000-0003-4592-4074}
\affiliation{%
\institution{Shandong University}
\city{Qingdao}
\country{China}
}
\email{chenzhumin@sdu.edu.cn}

\author{Zhaochun Ren}
\orcid{0000-0002-9076-6565}
\affiliation{%
\institution{Shandong University}
\city{Qingdao}
\country{China}
}
\email{zhaochun.ren@sdu.edu.cn}

\author{Chenyang Wang}
\orcid{0000-0001-6853-2116}
\affiliation{%
\institution{Shandong University}
\city{Qingdao}
\country{China}
}
\email{201900122032@mail.sdu.edu.cn}

\author{Qiang Yan}
\orcid{0000-0002-7328-2278}
\affiliation{%
\institution{WeChat, Tencent}
\city{Guangzhou}
\country{China}
}
\email{rolanyan@tencent.com}

\author{Maarten de Rijke}
\orcid{0000-0002-1086-0202}
\affiliation{%
\institution{University of Amsterdam}
\city{Amsterdam}
\country{The Netherlands}
}
\email{m.derijke@uva.nl}

\author{Xiuzhen Cheng}
\orcid{0000-0001-9998-2398}
\affiliation{%
\institution{Shandong University}
\city{Qingdao}
\country{China}
}
\email{xzcheng@sdu.edu.cn}

\author{Pengjie Ren}
\orcid{0000-0003-2964-6422}
\affiliation{%
\institution{Shandong University}
\city{Qingdao}
\country{China}
}
\email{renpengjie@sdu.edu.cn}
\authornote{Corresponding author.}

\renewcommand{\shortauthors}{Lin et al.}

\begin{document}

\title{Modeling Sequential Recommendation as Missing Information Imputation}

\begin{abstract}
Side information is being used extensively to improve the effectiveness of sequential recommendation models. 
It is said to help capture the transition patterns among items.
Most previous work on sequential recommendation that uses side information models item IDs and side information separately, which may fail to fully model the relation between the items and their side information. 
Moreover, in real-world systems, not all values of item feature fields are available.
This hurts the performance of models that rely on side information.
Existing methods tend to neglect the context of missing item feature fields, and fill them with generic or special values, e.g., \textit{unknown}, which might lead to sub-optimal performance.

To address the limitation of sequential recommenders with side information, we define a way to fuse side information and alleviate the problem of missing side information by proposing a unified task, namely the \acfi{MII}, which randomly masks some feature fields in a given sequence of items, including item IDs, and then forces a predictive model to recover them.
By considering the next item as a missing feature field, sequential recommendation can be formulated as a special case of \ac{MII}.
We propose a sequential recommendation model, called \acfi{MIIR}, that builds on the idea of \ac{MII} and simultaneously imputes missing item feature values and predicts the next item.
We devise a \acfi{DFSA} mechanism for \ac{MIIR} to capture all pairwise relations between items and their side information.
Empirical studies on three benchmark datasets demonstrate that \ac{MIIR}, supervised by \ac{MII}, achieves a significantly better sequential recommendation performance than state-of-the-art baselines.
\end{abstract}

\ccsdesc[500]{Information systems~Recommender systems}

\keywords{Sequential Recommendation, Side Information Fusion, Missing Information Imputation}


\maketitle


\section{Introduction}
Sequential recommendation models transition patterns among items and generates a recommendation for the next item~\citep{fang2020deep}. 
Traditional sequential recommendation solutions use the item ID as the only item feature field~\cite{BalzsHidasi2016SessionbasedRW,li2017neural,hidasi2018recurrent,tang2018personalized,kang2018self,wu2019session,sun2019bert4rec}.
In real-world cases, however, there is rich side information in the form of multiple types of structural feature fields, such as categories and brands, and unstructured feature fields, e.g., titles and descriptions, that can help to better model transitions between items.
In recent years, several publications have exploited side information to improve sequential recommendation performance~\cite{hidasi2016parallel,zhang2019feature,wang2020kerl,de2021transformers4rec,cai2021category,singer2022sequential,xie2022decoupled}.
Most focus on designing different mechanisms to fuse side information into recommendation models.
For example, \citet{hidasi2016parallel} use parallel \acp{RNN} \cite{lipton2015critical} to encode the information in item IDs and attributes, respectively,  and then combine the outputs of \acp{RNN} for item recommendation.
\citet{zhang2019feature} employ two groups of self-attention blocks~\citep{vaswani2017attention} for modeling items and features, and fuse them in the final stage.

Importantly, previous work for sequential recommendation with side information usually regards side information as an auxiliary representation of the item, so models item IDs and side information separately.
As a result, such methods only encode \textit{partial} relations in item sequences, e.g., the relation between an item and its side information, while the relation between an item and the side information of other items in the sequence is not well captured.

\begin{figure}[htbp]
    \centering
    \subfloat[Original sequence.]{\includegraphics[width=0.8\linewidth]{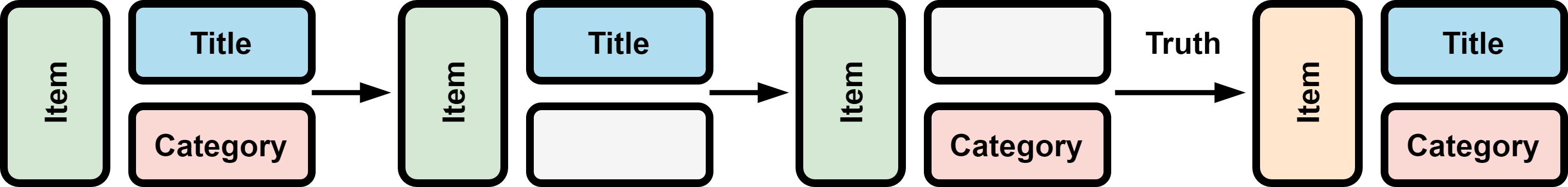}}
    \hfil
    \subfloat[Existing work without imputation.]{\includegraphics[width=0.8\linewidth]{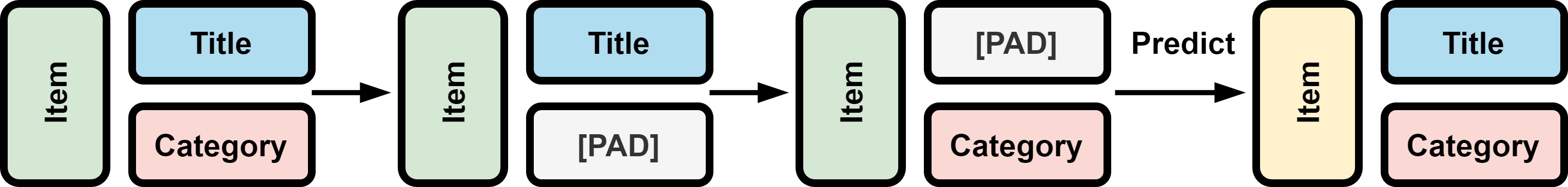}}
    \hfil
    \subfloat[Our work with imputation.]{\includegraphics[width=0.8\linewidth]{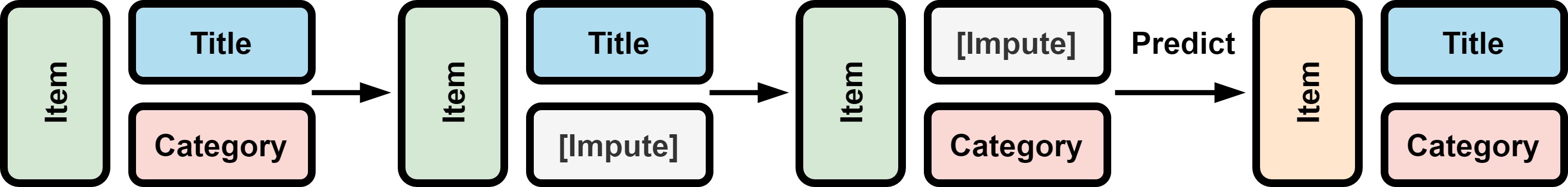}}
    \caption{Sequential recommendation of items with side information. Gray blocks represent missing information. ``[PAD]'' (in (b)) indicates padding with generic or special values as often done in existing work. ``[Impute]'' (in (c)) indicates imputation with actual values for missing feature fields.}    
    \label{figure_1_1}
\end{figure}

Even more importantly, previous studies often assume that all side information is available, which is rarely the case in real-world scenarios.
As illustrated in Fig.~\ref{figure_1_1}(a), i.e., the second and third items lack category and title information, respectively.
Previous work has proposed to fill such gaps with special values, such as a general category and a \textit{padding} text, to make models trainable and produce outputs.
However, for different items and item sequences, these special values are the same: they do not provide useful and specific information for recommendations and might introduce biases into the model learning instead~\citep{shi2019adaptive}.
As a result, as illustrated in Fig.~\ref{figure_1_1}(b), a model might recommend the wrong item.
Instead, we propose to \emph{impute} the missing side information, so that the recommendation model can use information from missing feature fields based on contexts, as illustrated in Fig.~\ref{figure_1_1}(c).

Some recent studies address the problem of missing side information in recommendation data.
\citet{wang2018lrmm} employ an \ac{AE} with a modality dropout to recover the missing rating and side information.
\citet{shi2019adaptive} propose an adaptive feature sampling strategy to introduce more missing feature fields into the training process, which increases the robustness of the recommendation model against missing side information.
\citet{wu2020joint} define item recommendation and attribute inference in a user-item bipartite graph with attributes, and propose a \ac{GCN} \cite{kipf2017semi} based model to join these two tasks.
However, the work just listed mainly targets non-sequential recommendation.
Moreover, it treats item recommendation and side information imputation as different tasks.

In this work, we seek to design a sequential recommendation model that can handle missing feature fields of items in items sequences.
The main challenge is how to adaptively impute missing information, including missing side information and the next item, according to the  information available in the item sequence.
First, we propose a task, the \acfi{MII} task
that randomly masks some non-missing feature fields, including item IDs, in the input sequence, and then asks the model to recover them in the output.
Since the next item to be recommended can also be seen as a missing feature field in the sequence, \ac{MII} unifies the missing side information imputation task with the next item prediction task.
\ac{MII} can be considered as an extension of the masked item prediction task \cite{zeng2021knowledge} that only masks item IDs. 
Based on the \ac{MII} task, we propose a sequential recommendation model, called \acfi{MIIR}, that jointly imputes missing side information and predicts the next item for the given item sequence.
\ac{MIIR} employs a \acfi{DFSA} mechanism to fuse the information in IDs and other feature fields for predicting both missing side information and the next item.
\ac{DFSA} captures the relation between any pair of feature fields in the input sequence, allowing it to fully fuse various types of (side) information to impute missing feature values and address the main recommendation challenge.

We conduct extensive experiments on three public datasets and show that \ac{MIIR} significantly outperforms state-of-the-art sequential recommendation baselines.
We also confirm that
\begin{enumerate*}[label=(\roman*)]
\item imputing missing side information and 
\item \ac{DFSA}
\end{enumerate*}
both help to improve the performance of sequential recommendation.

The main contributions of this work are as follows:
\begin{itemize}[leftmargin=*,nosep]
    \item We propose to unify the missing side information imputation task and the sequential recommendation task through \acf{MII}. To the best of our knowledge, this is the first work of its kind in sequential recommendation.
    \item We present a novel sequential recommendation model, \acf{MIIR}, that employs \ac{MII} to provide the signal for simultaneously imputing the missing item side information and predicting the next item and \acf{DFSA} to fuse various information.
    \item We conduct extensive experiments on three public datasets to verify the effectiveness of \ac{MII}, \ac{MIIR}, and \ac{DFSA} in sequential recommendation.
\end{itemize}

\section{Related Work}
In this section, we provide a review of research into sequential recommendation with side information, and research into missing side information in recommendation.

\subsection{Sequential recommendation with side information}
Side information fusion has been widely used in sequential recommendation because it can help to capture transition patterns among items.
We classify existing work into work that uses self-attention and work that does not.

As to work that does \emph{not} use self-attention, \citet{hidasi2016parallel} employ parallel \acp{RNN} to extract the information from ID sequences of item IDs and  sequences of features; they then examine different ways of combining the outputs of the \acp{RNN}.
\citet{zhou2020s3} propose self-supervised tasks to maximize the mutual information between an item and its attributes or between a sequence of item IDs and the sequence of their attributes.
\citet{yuan2021icai} construct a heterogeneous graph to aggregate different types of categorical attributes, then aggregate the representations of attribute types to get item representations.

Inspired by the success of self-attention mechanisms \cite{huang2018csan,tang2018self,zhao2020exploring}, some work uses self-attention to fuse items and side information.
\citet{zhang2019feature} first use a vanilla attention mechanism to fuse different types of side information on each item, and then use two branches of self-attention blocks to model transition patterns between IDs and side information; they then concatenate the hidden states of the two blocks for item recommendation.
\citet{liu2021noninvasive} propose a non-invasive self-attention mechanism that uses pure item ID representations as values and representations that integrate side information  as queries and keys to calculate the attention.
\citet{xie2022decoupled} decouple the non-invasive self-attention of different types of side information to get fused attention matrices for items.

Although many methods have been proposed for sequential recommendation with side information, they 
\begin{enumerate*}[label=(\roman*)]
\item neglect the missing information problem, and use fixed special values to fill missing feature fields, which might harm the performance, and 
\item hardly explore the relation between an item and the side information of other items in the same sequence.
\end{enumerate*}
These are aspects that we contribute on top of prior work.

\subsection{Missing side information in recommendation}
In real-world applications, the side information of users and items may be incomplete or missing, which may hurt the performance of recommendation models that rely on side information.

The traditional way to solve the problem of missing side information is to fill the missing feature fields with  heuristic values \cite{lee2018impute,biessmann2018deep,shi2019adaptive}, such as the most frequent feature values, average values, randomized values, the value \textit{unknown}, or  \textit{padding}.
As some studies have reported, these special values are independent of the context, and using them may lead to biased parameter estimation and prediction \cite{marlin2009collaborative,hernandez2014probabilistic}.
Another way to deal with missing feature fields is to impute their missing values.
Early approaches use KNN-based methods \cite{pan2015missing} or \acfp{AE} \cite{beaulieu2017missing,pereira2020reviewing} to predict the missing data.
\citet{wang2018lrmm} propose an \ac{AE}-based model with  modality dropout, which randomly drops representations of user or item information of different modalities in hidden states and reconstructs them by an \ac{AE}.
\citet{cao2019unifying} present a translation-based recommendation model that models preferences as translations from users to items, and jointly trains it with a \ac{KG} completion model that predicts the missing relations in the \ac{KG} for incorporating knowledge into the recommendation model. 
Instead of imputing the missing side information, \citet{shi2019adaptive} propose an adaptive feature sampling strategy, which employs layer-wise relevance propagation \cite{AlexanderBinder2016LayerWiseRP} to calculate the importance of different features and samples features to make the model more robust against unknown features.
\citet{wu2020joint} propose a \ac{GCN}-based model to jointly predict users’ preferences to items and predict the missing attribute values of users or items.

What we add on top of prior work on missing information in recommendation is that we focus on missing information in the context of \emph{sequential} recommendation.


\section{Method}
\subsection{Overview}
Before going into details of the proposed \ac{MII} task and \ac{MIIR} model, we introduce notation used in this paper.
We denote the item set as $I=\{\mathbf{i}_1,\ldots,\mathbf{i}_{N_i}\}$, where $N_i$ is the number of items and each item ID $\mathbf{i}_k \in \mathbb{R}^{N_i}$ is represented as a one-hot vector.
In addition to IDs, items have other feature fields corresponding to their side information.
In this work, we consider categorical feature fields, including category and brand, and textual feature fields, including title and description.
We denote the category set as $C=\{\mathbf{c}_1,\ldots,\mathbf{c}_{N_c}\}$, where $N_c$ is the number of categories and each category $\mathbf{c}_k \in \mathbb{R}^{N_c}$ is a one-hot vector.
Similarly, we denote the brand set as $B=\{\mathbf{b}_1,\ldots,\mathbf{b}_{N_b}\}$, where $N_b$ is the number of brands and each brand $\mathbf{b}_k \in \mathbb{R}^{N_b}$.
For titles and descriptions of items, we employ BERT \cite{devlin2019bert} to encode them into fixed-length vectors of size $768$.
We denote all titles and all descriptions as $T=\{\mathbf{t}_1,\ldots,\mathbf{t}_{N_i}\}$ and $D=\{\mathbf{d}_1,\ldots,\mathbf{d}_{N_i}\}$, respectively, where $\mathbf{t}_k$ and $\mathbf{d}_k \in \mathbb{R}^{768}$.
We use $S=[s_1,\ldots,s_n]$ to denote a sequence with $n$ items, where $s_k=[\mathbf{s}^i_k,\mathbf{s}^c_k,\mathbf{s}^b_k,\mathbf{s}^t_k,\mathbf{s}^d_k]$ is the sequence of features fields of the $k$-th item, $\mathbf{s}^i_k \in I$, $\mathbf{s}^c_k \subseteq C$, $\mathbf{s}^b_k \in B$, $\mathbf{s}^t_k \in T$, and $\mathbf{s}^d_k \in D$.
As an item may have multiple categories, we let $\mathbf{s}^c_k$ be a subset of $C$, which can be represented as a multi-hot vector $\mathbf{s}^c_k \in \mathbb{R}^{N_c}$.
For missing item IDs, categories and brands, we have special one-hot vectors denoted as $\mathbf{i}_{miss} \in I$, $\mathbf{c}_{miss} \in C$ and $\mathbf{b}_{miss} \in B$, respectively.
For missing titles and descriptions, we use the vector of ``[CLS][SEP]'' encoded by BERT to represent them, which are denoted as $\mathbf{t}_{miss} \in T$ and $\mathbf{d}_{miss} \in D$, respectively.
These missing representations will be used in both \ac{MIIR} and the baselines.
It is worth noting that other feature fields can be formalized and modeled in a similar way.

\begin{figure}[htbp]
    \centering    
    \subfloat[Sequential recommendation task.]{\includegraphics[width=0.65\linewidth]{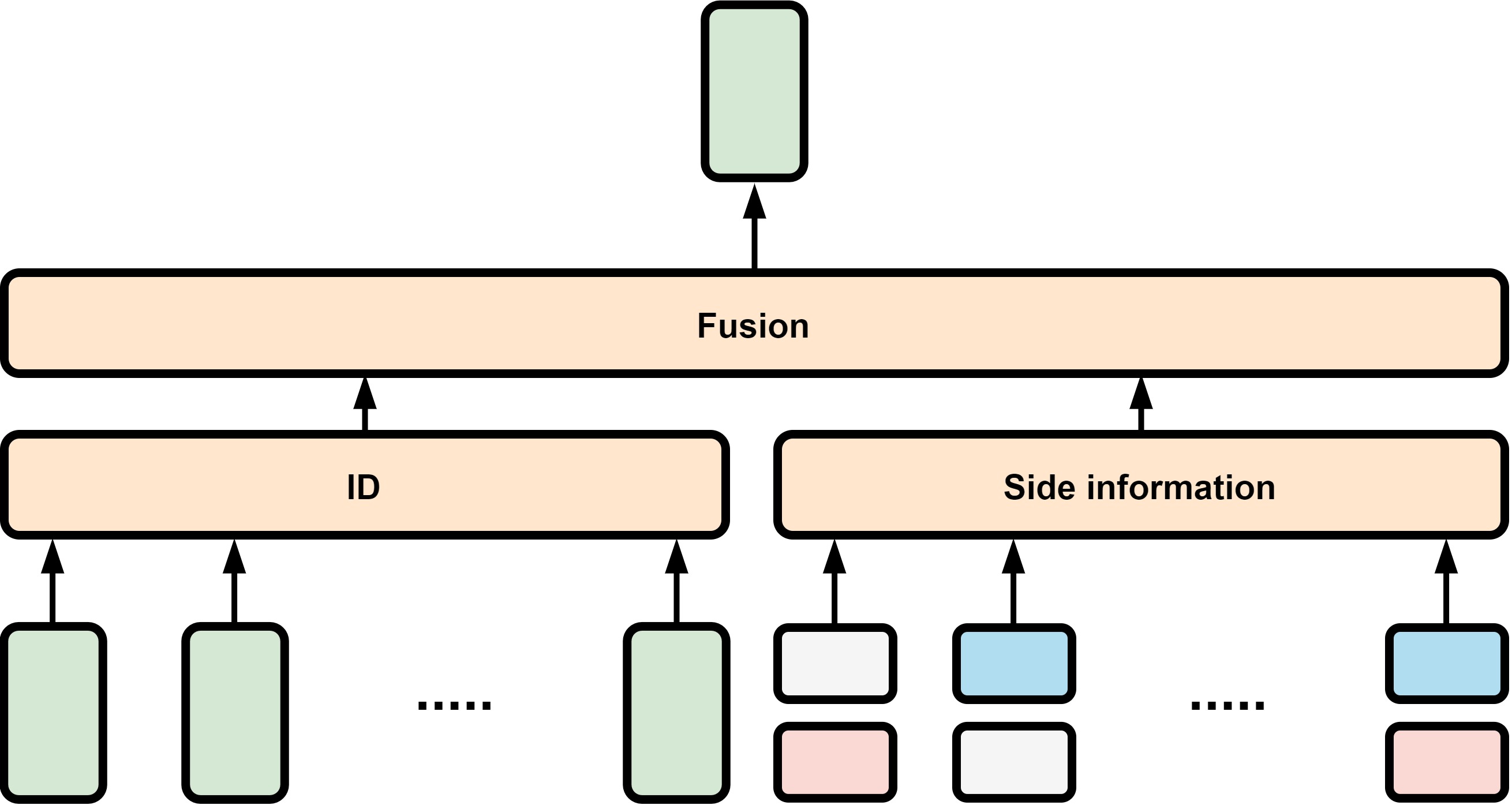}}
    \hfil
    \subfloat[Missing information imputation task.]{\includegraphics[width=0.65\linewidth]{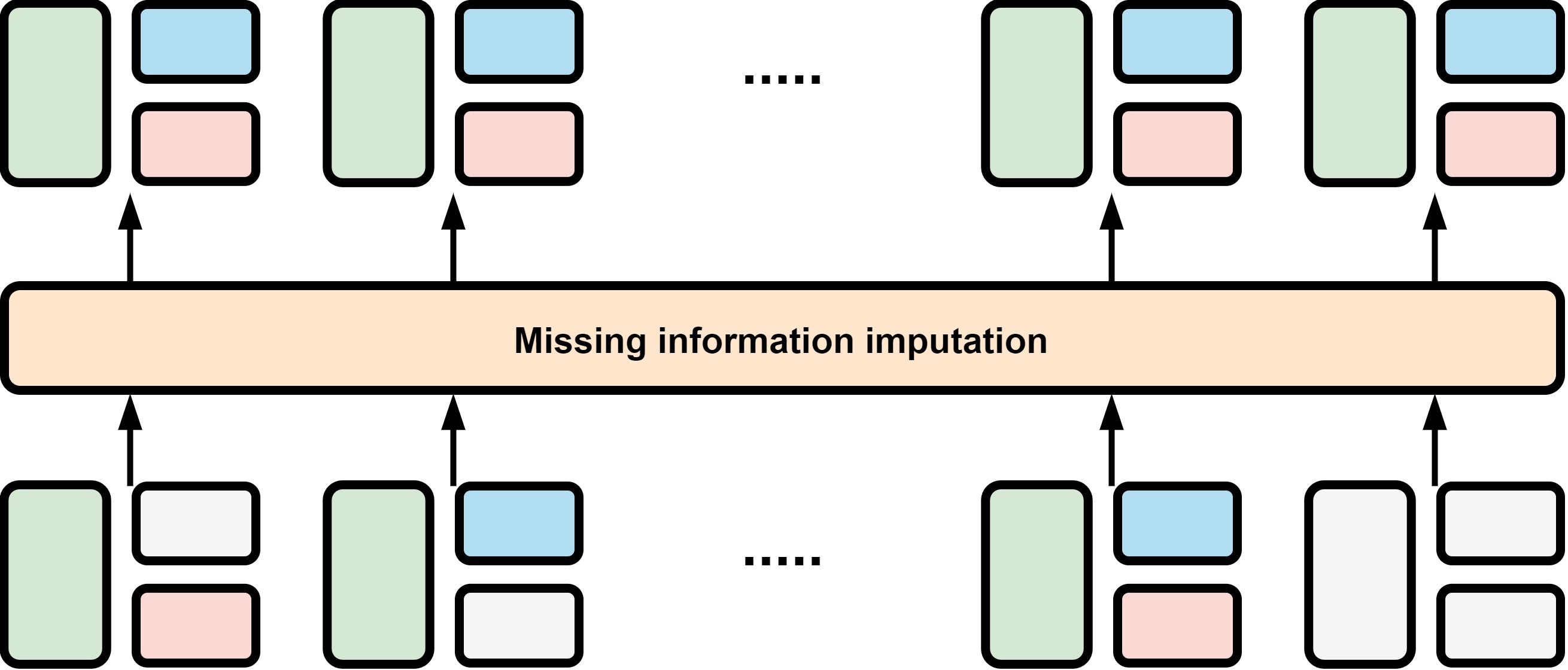}}    
    \caption{Comparing the sequential recommendation task and the missing information imputation task. (Same visual conventions as in Fig.~\ref{figure_1_1}.)}
    \label{figure_3_1}
\end{figure}

The \emph{missing information imputation} task is to impute the values of the missing feature fields in $S$.
The \emph{sequential recommendation} task is to predict the next item $s_{n+1}$ for $S$.
By appending a new item $s_{n+1}=[\mathbf{i}_{miss},\mathbf{c}_{miss},\mathbf{b}_{miss},\mathbf{t}_{miss},\mathbf{d}_{miss}]$ to the end of $S$ and imputing the $\mathbf{i}_{miss}$ of $s_{n+1}$, we can formulate the next item prediction task as a special case of missing information imputation task.
In Fig.~\ref{figure_3_1}, we compare the sequential recommendation task and the missing information imputation task.
In the sequential recommendation task, the next item is not considered as a missing data.
In the missing information imputation task, the next item is simply a missing feature field.
A model for the missing information imputation task that follows a unified way to impute both the next item and the other missing side information can be used for sequential recommendation.

\begin{figure}[htbp]
    \centering
    \includegraphics[width=0.8\linewidth]{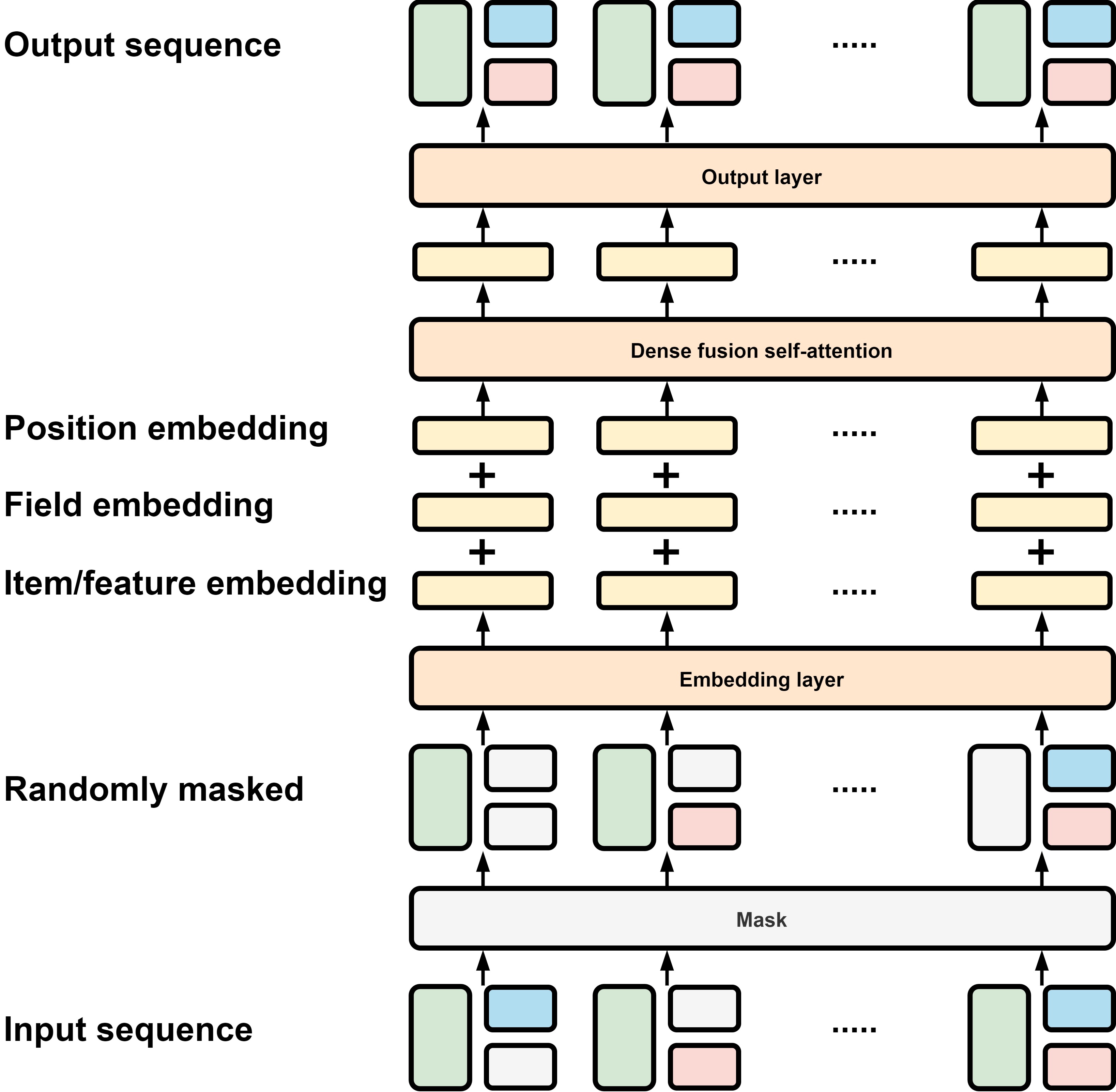}
    \vspace*{1mm}
    \caption{Architecture of the \acf{MIIR}. \ac{MIIR} takes a sequence of randomly masked feature fields as input. It transforms the input sequence into embeddings using the embedding layer. Then it employs a \acl{DFSA} mechanism to fuse information in the sequence. Finally, \ac{MIIR} uses an output layer to reconstruct the input sequence and calculate the \ac{MII} loss on masked feature fields. (Same visual conventions as in Fig.~\ref{figure_1_1}.)}
    \label{figure_3_2}
\end{figure}

To unify the missing side information imputation and next item recommendation tasks, we propose a sequential recommendation model called \acfi{MIIR}.
As we illustrate in Fig.~\ref{figure_3_2}, \ac{MIIR} consists of three main components:
\begin{enumerate*}[label=(\roman*)]
\item an embedding layer, 
\item a \acf{DFSA} mechanisms, and 
\item an output layer.
\end{enumerate*}
First, the embedding layer translates the input sequence into a series of embeddings.
Then, the \ac{DFSA} mechanism employs several transformer \cite{vaswani2017attention} layers to model the relation between any pair of feature fields in the sequence and fuse side information into the model for both imputation and recommendation.
Finally, the output layer imputes the missing feature values including item IDs in the sequence based on the output of \ac{DFSA}.
Next, we will introduce the details of these main components.

\subsection{Embedding layer}
The embedding layer projects all item feature fields in the input sequence into low-dimensional dense vectors with a unified length.

For the $k$-th item $s_k=[\mathbf{s}^i_k,\mathbf{s}^c_k,\mathbf{s}^b_k,\mathbf{s}^t_k,\mathbf{s}^d_k]$ in the given sequence $S$, the embedding layer uses different ways to translate different feature fields.
For the high-dimensional sparse vectors of $\mathbf{s}^i_k$, $\mathbf{s}^c_k$ and $\mathbf{s}^b_k$, we follow Eq.~\ref{e_k_1} to get the item embedding $\mathbf{e}^i_k \in \mathbb{R}^e$, the category embedding $\mathbf{e}^c_k \in \mathbb{R}^e$, and the brand embedding $\mathbf{e}^b_k \in \mathbb{R}^e$:
\begin{equation}
\begin{split}
\label{e_k_1}
\mathbf{e}^i_k & = \mathbf{E}^i\mathbf{s}^i_k, \\
\mathbf{e}^c_k & = \mathbf{E}^c\mathbf{s}^c_k, \\
\mathbf{e}^b_k & = \mathbf{E}^b\mathbf{s}^b_k,
\end{split}
\end{equation}
where $\mathbf{E}^i \in \mathbb{R}^{e\times{N_i}}$ is the item embedding matrix, $\mathbf{E}^c \in \mathbb{R}^{e\times{N_c}}$ is the category embedding matrix, $\mathbf{E}^b \in \mathbb{R}^{e\times{N_b}}$ is the brand embedding matrix, and $e$ is the embedding size.
For the high-dimensional dense vectors of $\mathbf{s}^t_k$ and $\mathbf{s}^d_k$, we project them into low-dimensional embeddings, i.e., the title embedding $\mathbf{e}^t_k \in \mathbb{R}^e$ and the description embedding $\mathbf{e}^d_k \in \mathbb{R}^e$, respectively, using Eq.~\ref{e_k_2}:
\begin{equation}
\begin{split}
\label{e_k_2}
\mathbf{e}^t_k & = \mathbf{E}^t\mathbf{s}^t_k, \\
\mathbf{e}^d_k & = \mathbf{E}^d\mathbf{s}^d_k,
\end{split}
\end{equation}
where $\mathbf{E}^t \in \mathbb{R}^{e\times{768}}$ and $\mathbf{E}^d \in \mathbb{R}^{e\times{768}}$ are the projection matrices.

In order to distinguish different types of feature fields for the same item, we learn a field embedding for each type of feature fields.
We denote the field embeddings of ID, category, brand, title and description as $\mathbf{f}^i$, $\mathbf{f}^c$, $\mathbf{f}^b$, $\mathbf{f}^t$ and $\mathbf{f}^d \in \mathbb{R}^e$, respectively.
To distinguish the items in different positions of the same sequence, we also inject the position information into the model by learning position embeddings, where the $k$-th position embedding is denoted as $\mathbf{p}_k \in \mathbb{R}^e$.
Finally, we add each field embedding to the corresponding item or feature embedding of $s_k$, and add $\mathbf{p}_k$ to all embeddings of $s_k$, as shown in Eq.~\ref{H_k}:
\begin{equation}
\label{H_k}
\mathbf{H}_k = 
\begin{bmatrix} 
\mathbf{h}^i_k \\
\mathbf{h}^c_k \\
\mathbf{h}^b_k \\
\mathbf{h}^t_k \\
\mathbf{h}^d_k
\end{bmatrix} = 
\begin{bmatrix}
\mathbf{e}^i_k+\mathbf{f}^i+\mathbf{p}_k \\
\mathbf{e}^c_k+\mathbf{f}^c+\mathbf{p}_k \\
\mathbf{e}^b_k+\mathbf{f}^b+\mathbf{p}_k \\
\mathbf{e}^t_k+\mathbf{f}^t+\mathbf{p}_k \\
\mathbf{e}^d_k+\mathbf{f}^d+\mathbf{p}_k
\end{bmatrix},
\end{equation}
where $\mathbf{h}^i_k$, $\mathbf{h}^c_k$, $\mathbf{h}^b_k$, $\mathbf{h}^t_k$, $\mathbf{h}^d_k \in \mathbb{R}^e$, and $\mathbf{H}_k \in \mathbb{R}^{5\times{e}}$ is the hidden state of $s_k$ that is the stack of all embeddings of its feature fields in order.

\subsection{Dense fusion self-attention}
The \acf{DFSA} mechanism follows a unified way to impute missing feature fields, both item IDs and side information.
To exploit the information in a given context for imputation, we need to model the relations between different feature fields and fuse the representations of various feature fields.
\ac{DFSA} calculates the attention values between any pair of feature fields and fuses the information of other feature fields based on the attention value.
By calculating the attention value, \ac{DFSA} captures all possible (hence \emph{dense}) pairwise relations between feature fields to facilitate missing information imputation.

Specifically, we first stack the hidden states of all items in $S$ in order by Eq.~\ref{H}:
\begin{equation}
\label{H}
\mathbf{H} = 
\begin{bmatrix} 
\mathbf{H}_1 \\  
\mathbf{H}_2 \\
\vdots \\
\mathbf{H}_n
\end{bmatrix},
\end{equation}
where $\mathbf{H} \in \mathbb{R}^{5n\times{e}}$ is the hidden state matrix of $S$.
Then, \ac{DFSA} employs a transformer with $L$ layers to update $\mathbf{H}$.

Each transformer layer $\mathrm{Trm}(\cdot)$ is composed of two sub-layers:
\begin{enumerate*}[label=(\roman*)]
\item multi-head self-attention $\operatorname{MH}(\cdot)$ and 
\item position-wise feed-forward $\operatorname{PFFN}(\cdot)$,
\end{enumerate*}
as defined in Eq.~\ref{transformer}:
\begin{equation}
\begin{split}
\label{transformer}
\mathbf{H}^{l+1} & = \mathrm{Trm}(\mathbf{H}^l) = \mathrm{LN}(\mathbf{\widetilde{H}}^l+\mathrm{Dropout}(\mathrm{PFFN}(\mathbf{\widetilde{H}}^l))) 
\\
\mathbf{\widetilde{H}}^l & = \mathrm{LN}(\mathbf{H}^l+\mathrm{Dropout}(\mathrm{MH}(\mathbf{H}^l))) 
\\
\mathrm{MH}(\mathbf{H}^l) & = [head_1;\ldots;head_h]\mathbf{W}^H 
\\
head_i & = \mathrm{Attn}(\mathbf{H}^l\mathbf{W}^Q_i,\mathbf{H}^l\mathbf{W}^K_i,\mathbf{H}^l\mathbf{W}^V_i) 
\\
\mathrm{Attn}(\mathbf{Q},\mathbf{K},\mathbf{V}) & = \mathrm{softmax}(\mathbf{Q}\mathbf{K}^{\top}/\sqrt{e}+\mathbf{M})\mathbf{V} 
\\
\mathrm{PFFN}(\mathbf{\widetilde{H}}^l) & = \mathrm{GELU}(\mathbf{\widetilde{H}}^l\mathbf{W}^F_1+\mathbf{b}^F_1)\mathbf{W}^F_2+\mathbf{b}^F_2,
\end{split}
\end{equation}
where $\mathrm{LN}$ is layer normalization  \cite{ba2016layer}, $\mathrm{Dropout}$ is dropout  \cite{srivastava2014dropout}, $\mathrm{Attn}$ is attention, $\mathrm{GELU}$ is a Gaussian error linear unit activation \cite{hendrycks2016gelu}, $[\ldots;\ldots]$ is the concatenation operation, $h$ is the number of heads, $\mathbf{W}^H \in \mathbb{R}^{e\times{e}}$, $\mathbf{W}^Q_i$, $\mathbf{W}^K_i$, $\mathbf{W}^V_i \in \mathbb{R}^{e\times{e/h}}$, $\mathbf{W}^F_1 \in \mathbb{R}^{e\times{4e}}$, $\mathbf{W}^F_2 \in \mathbb{R}^{4e\times{e}}$, $\mathbf{b}^F_1 \in \mathbb{R}^{4e}$ and $\mathbf{b}^F_2 \in \mathbb{R}^e$ are trainable parameters, $\mathbf{H}^l$ and $\mathbf{H}^{l+1} \in \mathbb{R}^{5n\times{e}}$ are the output hidden state matrices in the $l$-th layer and the $(l+1)$-th layer, and $\mathbf{H}^0=\mathbf{H}$.

The matrix $\mathbf{M} \in \mathbb{R}^{5n\times{5n}}$ in Eq.~\ref{transformer} is the attention mask which is defined as:
\begin{equation}
\label{M}
\mathbf{M}^{j,y}_{i,x} = \left\lbrace
\begin{array}{ll}
0, & \textrm{allow to attend,} \\
-\infty, & \textrm{prevent from attending,}
\end{array} \right.
\end{equation}
where $i$ and $j \in \{1,\ldots,n\}$, $x$ and $y \in \{i,c,b,t,d\}$, $\mathbf{M}^{j,y}_{i,x} \in \mathbf{M}$ is the mask to control whether the feature field $\mathbf{s}^y_j$ can attend to the feature field $\mathbf{s}^x_i$.
We set all $\mathbf{M}^{j,y}_{i,x}=0$,\footnote{Here we neglect the padding items.} which means we allow to attend between any pair of feature fields in the sequence.
Therefore, the \ac{DFSA} can model relations and fuse information between all possible pairs of feature fields to facilitate both imputation and recommendation.

\subsection{Output layer}
The output layer reconstructs the input feature fields based on the output hidden states of \ac{DFSA}.
First, we split the final output hidden state matrix $\mathbf{H}^L$ of \ac{DFSA} by Eq.~\ref{E}:
\begin{equation}
\label{E}
\mathbf{H}^L = \widehat{\mathbf{E}} = 
\begin{bmatrix} 
\widehat{\mathbf{E}}_1 \\
\widehat{\mathbf{E}}_2 \\
\vdots \\
\widehat{\mathbf{E}}_n
\end{bmatrix}, \quad \textrm{where }
\widehat{\mathbf{E}}_k = 
\begin{bmatrix} 
\hat{\mathbf{e}}^i_k \\
\hat{\mathbf{e}}^c_k \\
\hat{\mathbf{e}}^b_k \\
\hat{\mathbf{e}}^t_k \\
\hat{\mathbf{e}}^d_k
\end{bmatrix},
\end{equation}
and $\hat{\mathbf{e}}^i_k$, $\hat{\mathbf{e}}^c_k$, $\hat{\mathbf{e}}^b_k$, $\hat{\mathbf{e}}^t_k$, $\hat{\mathbf{e}}^d_k \in \mathbb{R}^e$.
Similar to the embedding layer, the output layer takes different ways to reconstruct different types of feature fields.
Specifically, for the categorical feature fields, we calculate the probability distributions $\mathbf{p}^i_k \in \mathbb{R}^{N_i}$, $\mathbf{p}^c_k \in \mathbb{R}^{N_c}$ and $\mathbf{p}^b_k \in \mathbb{R}^{N_b}$ of the item ID, category and brand of the $k$-th item $s_k$ as follows:
\begin{equation}
\begin{split}
\label{p}
\mathbf{p}^i_k & = \mathrm{softmax}({\mathbf{E}^i}^{\top}\hat{\mathbf{e}}^i_k) \\
\mathbf{p}^c_k & = \mathrm{sigmoid}({\mathbf{E}^c}^{\top}\hat{\mathbf{e}}^c_k) \\
\mathbf{p}^b_k & = \mathrm{softmax}({\mathbf{E}^b}^{\top}\hat{\mathbf{e}}^b_k),
\end{split}
\end{equation}
where $\mathbf{E}^i \in \mathbb{R}^{e\times{N_i}}$, $\mathbf{E}^c \in \mathbb{R}^{e\times{N_c}}$, $\mathbf{E}^b \in \mathbb{R}^{e\times{N_b}}$ are the re-used item embedding matrix, category embedding matrix, and brand embedding matrix in the embedding layer, respectively.
Note that we regard each category prediction as binary classification, because an item may contain multiple categories.
Then we obtain the reconstructed item ID $\hat{\mathbf{s}}^i_k \in \mathbb{R}^{N_i}$, category $\hat{\mathbf{s}}^c_k \in \mathbb{R}^{N_c}$ and brand $\hat{\mathbf{s}}^b_k \in \mathbb{R}^{N_b}$ based on the probability distributions, as shown in Eq.~\ref{s_k_1}:
\begin{equation}
\begin{split}
\label{s_k_1}
\hat{\mathbf{s}}^i_k & = \mathrm{argmax}(\mathbf{p}^i_k) \\
\hat{\mathbf{s}}^c_k & = \mathbf{1}(\mathbf{p}^c_k > 0.5) \\
\hat{\mathbf{s}}^b_k & = \mathrm{argmax}(\mathbf{p}^b_k),
\end{split}
\end{equation}
where $\mathbf{1}(\alpha)$ is the indicator function that equals 1 if $\alpha$ is true and 0 otherwise.
Meanwhile, for the textual feature fields, we follow Eq.~\ref{s_k_2} to get the reconstructed title \smash{$\hat{\mathbf{s}}^t_k \in \mathbb{R}^{768}$} and description \smash{$\hat{\mathbf{s}}^d_k \in \mathbb{R}^{768}$} directly:
\begin{equation}
\begin{split}
\label{s_k_2}
\hat{\mathbf{s}}^t_k & = \mathbf{O}^t\hat{\mathbf{e}}^t_k, \\
\hat{\mathbf{s}}^d_k & = \mathbf{O}^d\hat{\mathbf{e}}^d_k,
\end{split}
\end{equation}
where $\mathbf{O}^t \in \mathbb{R}^{768\times{e}}$ and $\mathbf{O}^d \in \mathbb{R}^{768\times{e}}$ are the projection matrices.

\subsection{Missing information imputation loss}
We train \ac{MIIR} with \ac{MII}.
\ac{MII} first randomly masks feature fields in the sequence with probability $p$, i.e., replacing a \emph{non-missing} feature value with the corresponding missing feature value  $\mathbf{i}_{miss}$, $\mathbf{c}_{miss}$, $\mathbf{b}_{miss}$, $\mathbf{t}_{miss}$ or $\mathbf{d}_{miss}$.
For the $k$-th item $s_k$ in the sequence $S$, we use $m^i_k$, $m^c_k$, $m^b_k$, $m^t_k$ and $m^d_k \in \{\mathit{true},\mathit{false}\}$ to denote whether its ID, category, brand, title and description are masked.
Then, \ac{MIIR} learns to recover the masked feature fields by \ac{MII} and impute the missing feature values based on the context.

Specifically, there are differences in the calculation of the missing information imputation loss for different types of feature fields.
For the categorical feature fields (i.e., ID, category and brand), our goal is to minimize the cross-entropy loss:
\begin{equation}
\begin{split}
\label{L^mii_k_1}
L^i_k & = -\mathbf{1}(m^i_k){\mathbf{s}^i_k}^{\top}\log(\mathbf{p}^i_k) 
\\
L^c_k & = -\mathbf{1}(m^c_k)({\mathbf{s}^c_k}^{\top}\log(\mathbf{p}^c_k)+(1-{\mathbf{s}^c_k}^{\top})\log(1-\mathbf{p}^c_k))/N_c 
\\
L^b_k & = -\mathbf{1}(m^b_k){\mathbf{s}^b_k}^{\top}\log(\mathbf{p}^b_k),
\end{split}
\end{equation}
where $L^i_k$, $L^c_k$ and $L^b_k$ are the imputation loss for the item ID, category and brand of $s_k$, respectively.
For the textual feature fields (i.e., title and description), our goal is to minimize the mean square error loss:
\begin{equation}
\begin{split}
\label{L^mii_k_2}
L^t_k & = \mathbf{1}(m^t_k)\|\mathbf{s}^t_k-\hat{\mathbf{s}}^t_k\|^2 \\
L^d_k & = \mathbf{1}(m^d_k)\|\mathbf{s}^d_k-\hat{\mathbf{s}}^d_k\|^2,
\end{split}
\end{equation}
where $L^t_k$ and $L^d_k$ are the imputation loss for the title and description of $s_k$.
The \emph{missing information imputation objective} of the entire model on $S$ is shown in Eq.~\ref{L^mii_S}:
\begin{equation}
\begin{split}
\label{L^mii_S}
L^{mii}_S & = 1/n\sum^n_{k=1} L^\mathit{mii}_k \\
L^{mii}_k & = L^i_k+L^c_k+L^b_k+L^t_k+L^d_k.
\end{split}
\end{equation}
Note that since the item ID is one of the feature fields and the next item prediction is a \ac{MII} task, \ac{MIIR} trained by \ac{MII} can directly be applied to sequential recommendation.

In our experiments, we also consider further fine-tuning \ac{MIIR} or directly training \ac{MIIR} with the masked item prediction loss to make the model only focus on the item prediction task.
Specifically, we randomly mask all feature fields of some items in the given sequence, while let \ac{MIIR} predict the masked item IDs only.
The \emph{recommendation loss} (i.e., the masked item prediction loss) on $S$ is defined as:
\begin{equation}
\begin{split}
\label{L^rec_S}
L^\mathit{rec}_S & = 1/n\sum^n_{k=1} L^\mathit{rec}_k 
\\
L^\mathit{rec}_k & = L^i_k = -\mathbf{1}(m^i_k){\mathbf{s}^i_k}^{\top}\mathrm{log}(\mathbf{p}^i_k),
\end{split}
\end{equation}
where $L^\mathit{rec}_k$ is the recommendation loss for $s_k$.

\section{Experimental Setup}
\subsection{Research questions}
In this paper, we seek to answer the following research questions:
\begin{enumerate}[label=\textbf{(RQ\arabic*)},leftmargin=*,nosep]
    \item How does \ac{MIIR} perform on the sequential recommendation task compared to state-of-the-art methods?
    \item What are the benefits of training \ac{MIIR} with \ac{MII}?
    \item Does modeling the relation between any pair of feature fields in item sequences help sequential recommendation?
    \item What is the performance of \ac{MIIR} on imputing missing side information?
\end{enumerate}

\subsection{Datasets}
\label{section:datasets}
There are many public datasets for experimenting with sequential recommendation; see \cite{fang2020deep}. 
However, we need sequential recommendation datasets that come with side information.
We conduct experiments on three public datasets: ``Beauty'', ``Sports and Outdoors'' and ``Toys and Games''~\cite{ni2019justifying}, as they have rich item side information, including category, brand, title and description.

We follow common practices \cite{zhang2019feature,liu2021noninvasive} to process the datasets.
We sort each user's records in chronological order to construct an item sequence.
We filter out item sequences whose length is less than $5$ to avoid noise from the cold-start problem.
For each item sequence, we use the last item for test, the second last item for validation, and the rest items for training.
For each test or validation item, we randomly sample $99$ negative items for ranking.
We randomly discard side information of items with probability $0.5$.
We use ``Beauty D'', ``Sports and Outdoors D'' and ``Toys and Games D'' to denote the datasets after discarding side information.
The statistics of the datasets after pre-processing are summarized in Table~\ref{dataset}.

\begin{table}[htbp]
\centering
\caption{Summary of the datasets. The missing rate is the percentage of missing feature fields in all feature fields. Especially, ``Missing rate D'' is the missing rate after discarding side information.}
\label{dataset}
\setlength{\tabcolsep}{1mm}
\begin{tabular}{l c c c}
\toprule
\bf Dataset & \bf Beauty & \bf Sports and Outdoors & \bf Toys and Games \\ 
\midrule
\#items & 121,291 & 194,715 & 164,978 \\
\#sequences & 52,374 & 84,368 & 58,314 \\
Average length & 8.97 & 8.50 & 8.99 \\
\#categories & 656 & 3,035 & 957 \\
\#brands & 13,188 & 14,163 & 14,135 \\
Missing rate & 12.54\% & 20.11\% & 11.20\% \\
Missing rate D & 56.32\% & 60.12\% & 55.51\% \\ 
\bottomrule
\end{tabular}
\end{table}

\subsection{Baselines}
\label{section:baselines}
We compare \ac{MIIR} with the following recommendation baselines, which can be grouped into
\begin{enumerate*}[label=(\roman*)]
\item methods without side information fusion, 
\item methods with side information fusion, and
\item methods with missing feature values.
\end{enumerate*}
\begin{itemize}[leftmargin=*]
\item \textbf{Methods without side information fusion:}
    \begin{itemize}
    \item \textbf{GRU4Rec} employs \acp{RNN} to capture sequential patterns between items for sequential recommendation~\cite{BalzsHidasi2016SessionbasedRW}.
    \item \textbf{SASRec} uses the self-attention mechanism to model item sequences for next item recommendations~\cite{kang2018self}.
    \item \textbf{BERT4Rec} uses a bidirectional self-attention network train-ed by a masked item prediction task for sequential recommendation~\cite{sun2019bert4rec}. 
    \end{itemize}
\item \textbf{Methods with side information fusion:}
    \begin{itemize}
    \item \textbf{PRNN} employs parallel \acp{RNN} to process items and their side information respectively, then combines the hidden states of the \acp{RNN} for next item prediction~\cite{hidasi2016parallel}.
    \item \textbf{FDSA} leverages two separate self-attention networks to model the ID transition patterns and the feature transition patterns respectively, then concatenates the outputs of two networks for next item prediction~\cite{zhang2019feature}.
    \item \textbf{NOVA} adopts a non-invasive self-attention mechanism to leverage side information under the BERT4Rec framework for sequential recommendation~\cite{liu2021noninvasive}.
    \end{itemize}
\item \textbf{Methods with missing feature values:}
    \begin{itemize}
    \item \textbf{RFS} randomly samples feature fields to introduce more missing information during training~\cite{shi2019adaptive}.
    RFS aims to make the model more robust with missing feature values instead of imputing missing feature fields.
    We combine RFS with FDSA and NOVA, and denote the variants as FDSA+RFS and NOVA+RFS.
    \item \textbf{LRMM} designs an auto-encoder with  modality dropout to impute both user ratings and missing side information for each item~\cite{wang2018lrmm}.
    LRMM is not proposed for sequential recommendation.
    Therefore, we use the imputed missing side information by LRMM to train FDSA and NOVA, and denote them as FDSA+LRMM and NOVA+LRMM.
    \end{itemize}
\end{itemize}
Other methods with side information fusion, such as \cite{zhou2020s3,yuan2021icai}, can only model categorical item side information; for a fair comparison, we do not consider them as baselines.
In addition to the baselines listed above, we compare \ac{MIIR} against four variants, namely \ac{MIIR}-F, \ac{MIIR}-R, \ac{MIIR}-M, and Sparse-\ac{MIIR}, to be defined in Section~\ref{RQ1}, ~\ref{RQ2} and ~\ref{RQ3}. 

We unify the sequential recommendation loss in all baselines, \ac{MIIR}, and its variants to the cross-entropy loss, rather than the pairwise loss \cite{rendle2009bpr}, to avoid noise due to negative sampling in the pairwise loss.

\subsection{Metrics and implementation}
To evaluate the performance of sequential recommendation methods, we employ two widely used evaluation metrics: HR@$k$ (hit ratio) and MRR (mean reciprocal rank) \cite{fang2020deep}, where $k \in \{5,10\}$.
\begin{itemize}[left=0em]
    \item \textbf{HR} measures the proportion of the sequences whose ground-truth items are amongst the top ranked items in all test sequences.
    \item \textbf{MRR} is the average of reciprocal ranks of the ground-truth items.
\end{itemize}

For all baselines and our proposed model, we initialize the trainable parameters randomly with the Xavier method \cite{glorot2010understanding}.
We train all methods with the Adam optimizer \cite{kingma2015adam} for $100$ epochs, with a batch size of $128$ and a learning rate of $0.0001$.
We also apply gradient clipping \cite{pascanu2013difficulty} with range $[-5,5]$ during training.
According to the average length in Table~\ref{dataset}, we set the maximum sequence length to $20$ for three datasets for all methods.

All hyper-parameters of the baselines are set following the suggestions from the original papers.
For the hyper-parameters of \ac{MIIR}, we set the embedding size $e$ to $64$, the number of heads $h$ to $4$, and the number of layers $L$ to $3$.
We set the dropout rate in \ac{DFSA} and the mask probability $p$ in \ac{MII} to $0.5$.

To facilitate reproducibility of the results reported in this paper, the code and data used in experiments are available at \url{https://github.com/TempSDU/MIIR}.


\section{Experimental Results}
\subsection{Overall performance}
\label{RQ1}

\begin{table}[htbp]
\centering
\caption{Performance comparison of \ac{MIIR}, variants, and the baselines on the ``Beauty'' dataset. \ac{MIIR}-F is a variant of \ac{MIIR} that is fine-tuned using the recommendation loss (see Section~\ref{RQ1}) and \ac{MIIR}-R is a variant trained using the recommendation loss only (see Section~\ref{RQ2}). The highest overall performance is denoted in bold face. The highest performance among the baselines is underlined. Impr.\ (\%) is the performance gain of \ac{MIIR} against the best baseline method. $*$ indicates that an improvement is statistically significant based on a two-sided paired t-test with $p<0.05$.}
\label{table_5_1}
\setlength{\tabcolsep}{1mm}
\begin{tabular}{l c c c c c c}
\toprule
                        & \multicolumn{3}{c}{\textbf{Beauty}}                       & \multicolumn{3}{c}{\textbf{Beauty D}}                     \\
\cmidrule(r){2-4} \cmidrule(r){5-7}
\textbf{Method}         & \textbf{HR@5}     & \textbf{HR@10}    & \textbf{MRR}      & \textbf{HR@5}     & \textbf{HR@10}    & \textbf{MRR}      \\
\midrule
GRU4Rec                 & 31.58             & 42.50             & 21.47             & 31.58             & 42.50             & 21.47             \\
SASRec                  & 32.83             & 43.61             & 23.16             & 32.83             & 43.61             & 23.16             \\
BERT4Rec                & 33.22             & 43.77             & 23.58             & 33.22             & 43.77             & 23.58             \\
\midrule
PRNN                    & 32.27             & 42.70             & 23.08             & 31.80             & 42.55             & 22.23             \\
FDSA                    & 35.22             & 44.83             & 25.39             & 35.02             & 44.68             & 25.33             \\
NOVA                    & 34.99             & 45.07             & 25.02             & 34.21             & 44.38             & 24.80             \\
\midrule
FDSA+RFS                & 35.45             & 45.40             & 25.68             & 34.73             & 44.56             & 25.17             \\
NOVA+RFS                & \underline{35.57} & \underline{45.61} & \underline{25.74} & 34.26             & 44.24             & 24.97             \\
LRMM                    & 22.74             & 32.95             & 17.09             & 18.04             & 26.94             & 13.96             \\
FDSA+LRMM               & 35.35             & 45.15             & 25.62             & \underline{35.10} & \underline{44.73} & \underline{25.52} \\
NOVA+LRMM               & 35.35             & 45.31             & 25.50             & 34.31             & 44.53             & 25.01             \\
\midrule
\ac{MIIR}               & \textbf{38.92}    & \textbf{48.61}    & \textbf{29.46}    & \textbf{37.30}    & \textbf{46.85}    & 27.90             \\
\ac{MIIR}-F             & 38.73             & 48.01             & 29.28             & 37.12             & 46.48             & \textbf{27.95}    \\
\ac{MIIR}-R             & 35.59             & 45.60             & 25.85             & 34.92             & 44.96             & 25.41             \\
\midrule
Impr. (\%)              & +3.35\rlap{$^*$}  & +3.00\rlap{$^*$}  & +3.72\rlap{$^*$}  & +2.20\rlap{$^*$}  & +2.12\rlap{$^*$}  & +2.38\rlap{$^*$}  \\
\bottomrule
\end{tabular}
\end{table}

\begin{table}[htbp]
\centering
\caption{Performance comparison of \ac{MIIR}, variants, and the baselines on the ``Sports and Outdoors'' dataset.}
\label{table_5_2}
\setlength{\tabcolsep}{1mm}
\begin{tabular}{l c c c c c c}
\toprule
                        & \multicolumn{3}{c}{\textbf{Sports and Outdoors}}          & \multicolumn{3}{c}{\textbf{Sports and Outdoors D}}        \\
\cmidrule(r){2-4} \cmidrule(r){5-7}
\textbf{Method}         & \textbf{HR@5}     & \textbf{HR@10}    & \textbf{MRR}      & \textbf{HR@5}     & \textbf{HR@10}    & \textbf{MRR}      \\
\midrule
GRU4Rec                 & 33.54             & 44.57             & 23.70             & 33.54             & 44.57             & 23.70             \\
SASRec                  & 34.46             & 44.69             & 25.41             & 34.46             & 44.69             & 25.41             \\
BERT4Rec                & 35.12             & 45.24             & 26.11             & 35.12             & 45.24             & 26.11             \\
\midrule
PRNN                    & 37.41             & 47.25             & 27.23             & 36.01             & 46.18             & 26.12             \\
FDSA                    & 39.16             & 48.08             & 29.27             & 37.30             & 46.74             & 27.20             \\
NOVA                    & 37.95             & 47.54             & 28.08             & 36.15             & 45.96             & 26.90             \\
\midrule
FDSA+RFS                & 38.18             & 47.18             & 28.31             & 37.17             & 46.65             & 27.01             \\
NOVA+RFS                & 37.63             & 47.41             & 27.33             & 35.86             & 45.52             & 26.84             \\
LRMM                    & 28.65             & 41.36             & 20.50             & 19.79             & 30.34             & 15.13             \\
FDSA+LRMM               & \underline{39.48} & \underline{48.52} & \underline{29.41} & \underline{38.46} & \underline{47.67} & \underline{28.24} \\
NOVA+LRMM               & 38.18             & 47.76             & 28.30             & 37.28             & 46.78             & 27.32             \\
\midrule
\ac{MIIR}               & \textbf{43.66}    & \textbf{52.63}    & \textbf{32.66}    & \textbf{40.55}    & \textbf{49.80}    & \textbf{30.04}    \\
\ac{MIIR}-F             & 42.66             & 51.49             & 32.01             & 39.98             & 48.98             & 29.86             \\
\ac{MIIR}-R             & 40.01             & 49.70             & 29.40             & 38.07             & 47.82             & 27.77             \\
\midrule
Impr. (\%)              & +4.18\rlap{$^*$}  & +4.11\rlap{$^*$}  & +3.25\rlap{$^*$}  & +2.09\rlap{$^*$}  & +2.13\rlap{$^*$}  & +1.80\rlap{$^*$}  \\
\bottomrule
\end{tabular}
\end{table}

\begin{table}[htbp]
\centering
\caption{Performance comparison of \ac{MIIR}, variants, and the baselines on the ``Toys and Games'' dataset.}
\label{table_5_3}
\setlength{\tabcolsep}{1mm}
\begin{tabular}{l c c c c c c}
\toprule
                        & \multicolumn{3}{c}{\textbf{Toys and Games}}               & \multicolumn{3}{c}{\textbf{Toys and Games D}}             \\
\cmidrule(r){2-4} \cmidrule(r){5-7}
\textbf{Method}         & \textbf{HR@5}     & \textbf{HR@10}    & \textbf{MRR}      & \textbf{HR@5}     & \textbf{HR@10}    & \textbf{MRR}      \\
\midrule
GRU4Rec                 & 31.19             & 42.15             & 21.90             & 31.19             & 42.15             & 21.90             \\
SASRec                  & 31.74             & 41.22             & 24.51             & 31.74             & 41.22             & 24.51             \\
BERT4Rec                & 31.45             & 41.22             & 23.25             & 31.45             & 41.22             & 23.25             \\
\midrule
PRNN                    & 34.00             & 44.25             & 24.32             & 32.71             & 42.98             & 23.23             \\
FDSA                    & 34.44             & 43.89             & 26.03             & 32.70             & 42.33             & 24.69             \\
NOVA                    & 34.50             & 44.34             & 25.86             & 34.00             & 43.74             & 25.06             \\
\midrule
FDSA+RFS                & 34.81             & 44.62             & 26.30             & 33.41             & 43.64             & 25.22             \\
NOVA+RFS                & 35.33             & 45.29             & 26.27             & 33.39             & 43.26             & 24.73             \\
LRMM                    & 29.88             & 40.96             & 21.87             & 19.85             & 29.83             & 15.15             \\
FDSA+LRMM               & 35.20             & 44.50             & 26.49             & 33.43             & 42.94             & 25.18             \\
NOVA+LRMM               & \underline{35.65} & \underline{45.50} & \underline{26.61} & \underline{34.51} & \underline{44.47} & \underline{25.51} \\
\midrule
\ac{MIIR}               & \textbf{40.11}    & \textbf{49.80}    & \textbf{29.64}    & \textbf{39.01}    & \textbf{48.89}    & 28.74             \\
\ac{MIIR}-F             & 39.00             & 47.76             & 29.57             & 38.25             & 47.45             & \textbf{28.75}    \\
\ac{MIIR}-R             & 35.80             & 45.37             & 26.00             & 34.69             & 44.30             & 24.81             \\
\midrule
Impr. (\%)              & +4.46\rlap{$^*$}  & +4.30\rlap{$^*$}  & +3.03\rlap{$^*$}  & +4.50\rlap{$^*$}  & +4.42\rlap{$^*$}  & +3.23\rlap{$^*$}  \\
\bottomrule
\end{tabular}
\end{table}

To answer RQ1, we compare \ac{MIIR} against the recommendation models listed in Section~\ref{section:baselines} on the three datasets from Section~\ref{section:datasets}.
Table~\ref{table_5_1}, \ref{table_5_2} and \ref{table_5_3} list the evaluation results of all methods on each dataset, respectively.
Based on these results, we have the following observations.

First, on all datasets, \ac{MIIR} performs significantly better than all baselines by a large margin despite the different missing rates, in terms of HR@5, HR@10 and MRR.
\ac{MIIR} has two major advantages:
\begin{enumerate*}[label=(\roman*)]
\item \ac{MIIR} trains the model using \ac{MII} to enhance its ability  to deal with missing side information in sequential recommendation (see detailed analysis in Section \ref{RQ2}), and
\item \ac{MIIR} employs \ac{DFSA} to improve the side information fusion in the model (see Section \ref{RQ3} for further analysis).
\end{enumerate*}

Second, the item side information can help sequential recommender systems to more accurately model the transition patterns among items.
To verify this, we divide all methods into three groups: 
\begin{enumerate*}[label=(\roman*)]
\item GRU4Rec and PRNN that are based on \acp{RNN}; 
\item SASRec and FDSA that are based on left-to-right self-attention networks; and
\item BERT4Rec, NOVA, and \ac{MIIR} that employ bidirectional self-attention networks and the masked item prediction task.
\end{enumerate*}
In each group, we see that methods that fuse side information outperform methods that only rely on item IDs, which illustrates that item side information does help.

Third, the performance of PRNN, FDSA, NOVA and \ac{MIIR} on the ``Beauty'', ``Sports and Outdoors'' and ``Toys and Games'' datasets is higher than that on the discarded versions of the datasets (i.e., ``Beauty D'', ``Sports and Outdoors D'' and ``Toys and Games D'').
We see two reasons for this difference:
\begin{enumerate*}[label=(\roman*)]
\item the ``Beauty D'', ``Sports and Outdoors D'' and ``Toys and Games D'' datasets discard some side information, so the available side information becomes less, and
\item using the special values (i.e., $\mathbf{i}_{miss}$, $\mathbf{c}_{miss}$, $\mathbf{b}_{miss}$, $\mathbf{t}_{miss}$ and $\mathbf{d}_{miss}$) to fill missing feature fields may be harmful to PRNN, FDSA and NOVA.
\end{enumerate*}

Fourth, by comparing FDSA+RFS and NOVA+RFS with FDSA and NOVA, we can see that RFS does not consistently improve the performance of FDSA and NOVA on all datasets.
RFS even degrades the performance of FDSA and NOVA in some cases.
Because RFS introduces more missing feature values into the model training instead of imputing missing feature fields, it does not deal with the missing side information problem fundamentally.

Fifth, the performance of LRMM is significantly worse than that of the sequential recommendation models with side information.
LRMM even performs worse than GRU4Rec, SASRec and BERT4Rec that neglect the item side information.
The main reason is that LRMM is not a sequential model, so it does not exploit the relation and information in sequences to make recommendation and imputation, both of which are essential in the sequential recommendation task.
We can also observe that FDSA+LRMM and NOVA+LRMM outperform FDSA and NOVA, which verifies the effectiveness of the imputation results of LRMM.
This also demonstrates that imputing missing feature values is a better way to alleviate the missing side information problem than using fixed special values and RFS.

Sixth, modeling sequential recommendation as \acl{MII} is sufficient to train a recommendation model.
To verify this, we conduct an experiment that first pre-trains \ac{MIIR} using the missing information imputation loss (Eq.~\ref{L^mii_S}), and then fine-tunes it using the recommendation loss (Eq.~\ref{L^rec_S}). We use \ac{MIIR}-F to denote this variant of \ac{MIIR}.
In Table~\ref{table_5_1} we see that \ac{MIIR}-F performs worse than \ac{MIIR} in most cases.
Fine-tuning \ac{MIIR}-F with the recommendation loss might lead to overfitting, resulting in performance decreases.
This result supports the conclusion that with \ac{MII} we can unify the sequential recommendation task as a particular type of missing information imputation task to train \ac{MIIR} together with the other imputation task for missing item side information.

\subsection{Benefits of \ac{MII}}
\label{RQ2}

\begin{table}[htbp]
\centering
\caption{Performance comparison to assess the use of missing feature fields on the ``Beauty'' dataset. \ac{MIIR}-M and \ac{MIIR}-R-M are the variants of \ac{MIIR} and \ac{MIIR}-R respectively that mask missing feature fields in self-attention (see Section~\ref{RQ2}).}
\label{table_5_4}
\setlength{\tabcolsep}{1mm}
\begin{tabular}{l c c c c c c}
\toprule
                        & \multicolumn{3}{c}{\textbf{Beauty}}                       & \multicolumn{3}{c}{\textbf{Beauty D}}                     \\
\cmidrule(r){2-4} \cmidrule(r){5-7}
\textbf{Method}         & \textbf{HR@5}     & \textbf{HR@10}    & \textbf{MRR}      & \textbf{HR@5}     & \textbf{HR@10}    & \textbf{MRR}      \\
\midrule
\ac{MIIR}               & 38.92             & 48.61             & \textbf{29.46}    & \textbf{37.30}    & \textbf{46.85}    & \textbf{27.90}    \\
\ac{MIIR}-R             & 35.59             & 45.60             & 25.85             & 34.92             & 44.96             & 25.41             \\
\midrule
\ac{MIIR}-M             & \textbf{39.16}    & \textbf{48.67}    & 29.45             & 37.12             & 46.58             & 27.83             \\
\ac{MIIR}-R-M           & 36.40             & 46.31             & 27.11             & 34.71             & 45.01             & 25.42             \\
\bottomrule
\end{tabular}
\end{table}

\begin{table}[htbp]
\centering
\caption{Performance comparison to assess the use of missing feature fields on the ``Sports and Outdoors'' dataset.}
\label{table_5_5}
\setlength{\tabcolsep}{1mm}
\begin{tabular}{l c c c c c c}
\toprule
                        & \multicolumn{3}{c}{\textbf{Sports and Outdoors}}          & \multicolumn{3}{c}{\textbf{Sports and Outdoors D}}        \\
\cmidrule(r){2-4} \cmidrule(r){5-7}
\textbf{Method}         & \textbf{HR@5}     & \textbf{HR@10}    & \textbf{MRR}      & \textbf{HR@5}     & \textbf{HR@10}    & \textbf{MRR}      \\
\midrule
\ac{MIIR}               & \textbf{43.66}    & \textbf{52.63}    & \textbf{32.66}    & \textbf{40.55}    & \textbf{49.80}    & \textbf{30.04}    \\
\ac{MIIR}-R             & 40.01             & 49.70             & 29.40             & 38.07             & 47.82             & 27.77             \\
\midrule
\ac{MIIR}-M             & 43.04             & 52.12             & 32.16             & 40.36             & 49.65             & 29.81             \\
\ac{MIIR}-R-M           & 39.71             & 48.98             & 29.15             & 38.33             & 48.10             & 28.12             \\
\bottomrule
\end{tabular}
\end{table}

\begin{table}[htbp]
\centering
\caption{Performance comparison to assess the use of missing feature fields on the ``Toys and Games'' dataset.}
\label{table_5_6}
\setlength{\tabcolsep}{1mm}
\begin{tabular}{l c c c c c c}
\toprule
                        & \multicolumn{3}{c}{\textbf{Toys and Games}}               & \multicolumn{3}{c}{\textbf{Toys and Games D}}             \\
\cmidrule(r){2-4} \cmidrule(r){5-7}
\textbf{Method}         & \textbf{HR@5}     & \textbf{HR@10}    & \textbf{MRR}      & \textbf{HR@5}     & \textbf{HR@10}    & \textbf{MRR}      \\
\midrule
\ac{MIIR}               & \textbf{40.11}    & \textbf{49.80}    & \textbf{29.64}    & \textbf{39.01}    & \textbf{48.89}    & \textbf{28.74}    \\
\ac{MIIR}-R             & 35.80             & 45.37             & 26.00             & 34.69             & 44.30             & 24.81             \\
\midrule
\ac{MIIR}-M             & 39.33             & 49.22             & 28.97             & 37.80             & 47.58             & 27.82             \\
\ac{MIIR}-R-M           & 35.22             & 45.29             & 26.28             & 34.53             & 44.47             & 25.58             \\
\bottomrule
\end{tabular}
\end{table}

To answer RQ2, we analyze how \ac{MIIR} benefits from training with \ac{MII}.

In Table~\ref{table_5_1}, \ref{table_5_2} and \ref{table_5_3}, we report on results of a variant of \ac{MIIR} that directly trains \ac{MIIR} with the recommendation loss shown in Eq.~\ref{L^rec_S}.
We write \ac{MIIR}-R for this variant of \ac{MIIR} without the supervised signal of \ac{MII}.
When we compare the performance of \ac{MIIR} and \ac{MIIR}-R, we see very substantial gaps.
This confirms the effectiveness of training \ac{MIIR} with \ac{MII}, which accounts for the main part of the improvement of \ac{MIIR} over other methods.

To demonstrate that \ac{MIIR} can mine useful information from missing feature fields by training with \ac{MII}, we design a variant of \ac{MIIR} called \ac{MIIR}-M by masking missing feature fields.
In \ac{MIIR}-M, we revise the attention mask $\mathbf{M}$ used in Eq.~\ref{transformer}, which is a null matrix in \ac{MIIR}.
The revision in $\mathbf{M}$ is defined as:
\begin{equation}
\label{revised_M}
\mathbf{M}^{j,y}_{i,x} = \left\lbrace
\begin{array}{ll}
-\infty, & \textrm{if $\mathbf{s}^x_i$ or $\mathbf{s}^y_j \in \{\mathbf{c}_{miss},\mathbf{b}_{miss},\mathbf{t}_{miss},\mathbf{d}_{miss}\}$,} \\
0, & \textrm{otherwise,}
\end{array} \right.
\end{equation}
where the condition of $\mathbf{s}^x_i$ or $\mathbf{s}^y_j \in \{\mathbf{c}_{miss}$, $\mathbf{b}_{miss}$, $\mathbf{t}_{miss}$, $\mathbf{d}_{miss}\}$ depends on the original input sequence instead of the sequence after randomly masking.
The purpose of the variant is to prevent the model from attending to the missing feature fields about item side information in the sequence.
On the one hand, \ac{MIIR}-M cannot mine and fuse any information in missing feature fields for sequential recommendation.
On the other hand, \ac{MIIR}-M is unable to exploit the information in non-missing feature fields to impute the missing side information.
Besides, we mask missing feature fields for \ac{MIIR}-R to analyze how missing feature values affect the performance of \ac{MIIR} without \ac{MII}, denoted as \ac{MIIR}-R-M.

In Table~\ref{table_5_4}, \ref{table_5_5} and \ref{table_5_6}, we compare \ac{MIIR} and \ac{MIIR}-R with \ac{MIIR}-M and \ac{MIIR}-R-M, respectively.
We find that \ac{MIIR} outperforms \ac{MIIR}-M in most cases, which indicates that \ac{MIIR} is able to extract useful information from missing feature fields to improve the sequential recommendation performance.
We also observe that \ac{MIIR}-R-M performs better than \ac{MIIR}-R in some cases.
This phenomenon indicates that using fixed special values for filling missing feature fields hurts the model performance.
Instead, masking missing feature fields is a better way without imputation.
On the ``Beauty'' dataset, \ac{MIIR} only achieves comparable performance with \ac{MIIR}-M, and \ac{MIIR}-R also performs worse than \ac{MIIR}-R-M.
However, the performance gap between \ac{MIIR} and \ac{MIIR}-M is smaller than that between \ac{MIIR}-R and \ac{MIIR}-R-M, and we have similar observations on other datasets.
This illustrates that imputing missing feature values is to be preferred over masking them for alleviating the missing side information problem.

\ac{MIIR}-M also outperforms all baselines on the three datasets with different missing rates.
Training \ac{MIIR} with \ac{MII} helps \ac{MIIR} to make use \emph{non-missing} feature fields.
Imputing the masked non-missing feature values requires the model to capture the relations between different feature fields, so \ac{MII} guides \ac{MIIR} to better fuse side information into the model for improving the sequential recommendation performance.

\subsection{Effectiveness of \ac{DFSA}}
\label{RQ3}

\begin{table}[htbp]
\centering
\caption{Performance comparison of dense and sparse attention on the ``Beauty'' dataset. Sparse-\ac{MIIR} and Sparse-\ac{MIIR}-R are variants of \ac{MIIR} and \ac{MIIR}-R, respectively, in which \ac{DFSA} is replaced by \acs{SFSA} (see Section~\ref{RQ3}).}
\label{table_5_7}
\setlength{\tabcolsep}{1mm}
\begin{tabular}{l c c c c c c}
\toprule
                        & \multicolumn{3}{c}{\textbf{Beauty}}                       & \multicolumn{3}{c}{\textbf{Beauty D}}                     \\
\cmidrule(r){2-4} \cmidrule(r){5-7}
\textbf{Method}         & \textbf{HR@5}     & \textbf{HR@10}    & \textbf{MRR}      & \textbf{HR@5}     & \textbf{HR@10}    & \textbf{MRR}      \\
\midrule
\ac{MIIR}               & \textbf{38.92}    & \textbf{48.61}    & \textbf{29.46}    & \textbf{37.30}    & \textbf{46.85}    & \textbf{27.90}    \\
\ac{MIIR}-R             & 35.59             & 45.60             & 25.85             & 34.92             & 44.96             & 25.41             \\
\midrule
Sparse-\ac{MIIR}        & 36.71             & 46.60             & 26.87             & 36.04             & 45.98             & 26.34             \\
Sparse-\ac{MIIR}-R      & 34.95             & 45.02             & 25.35             & 34.61             & 44.84             & 25.19             \\
\bottomrule
\end{tabular}
\end{table}

\begin{table}[htbp]
\centering
\caption{Performance comparison of dense and sparse attention on the ``Sports and Outdoors'' dataset.}
\label{table_5_8}
\setlength{\tabcolsep}{1mm}
\begin{tabular}{l c c c c c c}
\toprule
                        & \multicolumn{3}{c}{\textbf{Sports and Outdoors}}          & \multicolumn{3}{c}{\textbf{Sports and Outdoors D}}        \\
\cmidrule(r){2-4} \cmidrule(r){5-7}
\textbf{Method}         & \textbf{HR@5}     & \textbf{HR@10}    & \textbf{MRR}      & \textbf{HR@5}     & \textbf{HR@10}    & \textbf{MRR}      \\
\midrule
\ac{MIIR}               & \textbf{43.66}    & \textbf{52.63}    & \textbf{32.66}    & \textbf{40.55}    & \textbf{49.80}    & \textbf{30.04}    \\
\ac{MIIR}-R             & 40.01             & 49.70             & 29.40             & 38.07             & 47.82             & 27.77             \\
\midrule
Sparse-\ac{MIIR}        & 40.52             & 50.04             & 29.64             & 39.24             & 48.91             & 28.67             \\
Sparse-\ac{MIIR}-R      & 38.61             & 48.29             & 28.25             & 37.56             & 47.72             & 27.21             \\
\bottomrule
\end{tabular}
\end{table}

\begin{table}[htbp]
\centering
\caption{Performance comparison of dense and sparse attention on the ``Toys and Games'' dataset.}
\label{table_5_9}
\setlength{\tabcolsep}{1mm}
\begin{tabular}{l c c c c c c}
\toprule
                        & \multicolumn{3}{c}{\textbf{Toys and Games}}               & \multicolumn{3}{c}{\textbf{Toys and Games D}}             \\
\cmidrule(r){2-4} \cmidrule(r){5-7}
\textbf{Method}         & \textbf{HR@5}     & \textbf{HR@10}    & \textbf{MRR}      & \textbf{HR@5}     & \textbf{HR@10}    & \textbf{MRR}      \\
\midrule
\ac{MIIR}               & \textbf{40.11}    & \textbf{49.80}    & \textbf{29.64}    & \textbf{39.01}    & \textbf{48.89}    & \textbf{28.74}    \\
\ac{MIIR}-R             & 35.80             & 45.37             & 26.00             & 34.69             & 44.30             & 24.81             \\
\midrule
Sparse-\ac{MIIR}        & 37.61             & 47.77             & 27.23             & 37.06             & 47.27             & 26.79             \\
Sparse-\ac{MIIR}-R      & 35.58             & 45.66             & 25.80             & 34.46             & 44.54             & 24.53             \\
\bottomrule
\end{tabular}
\end{table}

\begin{table}[htbp]
\centering
\caption{Performance comparison of LRMM and \ac{MIIR} for missing side information imputation on the ``Beauty D'', ``Sports and Outdoors D'' and ``Toys and Games D'' datasets.}
\label{table_5_10}
\setlength{\tabcolsep}{1mm}
\begin{tabular}{l l l cc}
\toprule
\textbf{Dataset} & \textbf{Field} & \textbf{Metric} & LRMM & \ac{MIIR}     
 \\
\midrule
\multirow{6}{*}{\textbf{Beauty D}} & \multirow{3}{*}{\textbf{Category}} & \textbf{Precision} & 70.15\phantom{00} & \textbf{79.64}\phantom{00} 
\\
& & \textbf{Recall} & \textbf{48.41}\phantom{00} & 36.97\phantom{00} 
\\
& & \textbf{F1} & \textbf{52.96}\phantom{00} & 48.61\phantom{00} 
\\
& \textbf{Brand} & \textbf{Accuracy} & \phantom{0}\textbf{7.84}\phantom{00}  & \phantom{0}5.01\phantom{00}
\\
& \textbf{Title} & \textbf{Mean squared error} & \phantom{0}0.0871 & \phantom{0}\textbf{0.0514} 
\\
& \textbf{Description} & \textbf{Mean squared error} & \phantom{0}0.1454 & \phantom{0}\textbf{0.0704} 
\\
\midrule
\multirow{6}{*}{\textbf{\begin{tabular}[c]{@{}l@{}}Sports and \\ Outdoors D\end{tabular}}} & \multirow{3}{*}{\textbf{Category}} & \textbf{Precision} & 57.02\phantom{00} & \textbf{74.97}\phantom{00}  
\\
 & & \textbf{Recall} & \textbf{51.31}\phantom{00} & 35.91\phantom{00}           
\\
&  & \textbf{F1} & 46.38\phantom{00} & \textbf{46.40}\phantom{00}  
\\
& \textbf{Brand} & \textbf{Accuracy} & \phantom{0}\textbf{6.06}\phantom{00}  & \phantom{0}4.43\phantom{00}            
\\
& \textbf{Title} & \textbf{Mean squared error} & \phantom{0}0.0927 & \phantom{0}\textbf{0.0534} 
\\
& \textbf{Description} & \textbf{Mean squared error} & \phantom{0}0.1474 & \phantom{0}\textbf{0.0835} 
\\
\midrule
\multirow{6}{*}{\textbf{\begin{tabular}[c]{@{}l@{}}Toys and \\ Games D\end{tabular}}} & \multirow{3}{*}{\textbf{Category}} & \textbf{Precision} & 72.32\phantom{00} & \textbf{89.31}\phantom{00}  
\\
& & \textbf{Recall} & \textbf{51.08}\phantom{00} & 42.31\phantom{00} 
\\
& & \textbf{F1} & 54.11\phantom{00} & \textbf{55.50}\phantom{00}  
\\
& \textbf{Brand} & \textbf{Accuracy} & \textbf{18.61}\phantom{00} & 14.49\phantom{00}           
\\
& \textbf{Title} & \textbf{Mean squared error} & \phantom{0}0.0858 & \phantom{0}\textbf{0.0514} 
\\
& \textbf{Description} & \textbf{Mean squared error} & \phantom{0}0.1427 & \phantom{0}\textbf{0.0777} 
\\
\bottomrule
\end{tabular}
\end{table}

To answer RQ3, we conduct an ablation study to analyze the effectiveness of \ac{DFSA} in \ac{MIIR}.

We first compare \ac{MIIR}-R, the variant of \ac{MIIR} that is trained with recommendation loss only, with the baselines in Table~\ref{table_5_1}, \ref{table_5_2}, \ref{table_5_3}.
\ac{MIIR}-R achieves better or comparable performance with the baselines on most evaluation metrics of all datasets, even without the help of \ac{MII}.
The main reason is that \ac{MIIR}-R has \acf{DFSA} to better fuse information in the item sequence for improving sequential recommendation.

In order to validate that it is important to model \emph{all} possible pairwise relations in an item sequence for sequential recommendation, we design another self-attention mechanism called \acfi{SFSA}.
\ac{SFSA} modifies the attention mask $\mathbf{M}$ in Eq.~\ref{transformer} into:
\begin{equation}
\label{sfsa}
\mathbf{M}^{j,y}_{i,x} = \left\lbrace
\begin{array}{ll}
0, & \textrm{if $i==j$ or $x==y$,} \\
-\infty, & \textrm{otherwise,}
\end{array} \right.
\end{equation}
where the condition $i==j$ or $x==y$ means that \ac{SFSA} only allows to attend between the pair of feature fields belonging to the same item or the same type.
Therefore, \ac{SFSA} only models the relation between different feature fields of the same item or the relation between the same type of feature fields of different items in the sequence.
These relations are also modeled in some baselines, such as PRNN and FDSA.

In Table~\ref{table_5_7}, \ref{table_5_8} and \ref{table_5_9}, we compare the performance of \ac{DFSA} and \ac{SFSA} as components of \ac{MIIR} and \ac{MIIR}-R.
We write Sparse-\ac{MIIR} and Sparse-\ac{MIIR}-R for the variants of \ac{MIIR} and \ac{MIIR}-R, respectively, in which \ac{DFSA} is replaced by \ac{SFSA}.
We can see that \ac{MIIR} outperforms Sparse-\ac{MIIR} on all datasets despite different missing rates.
What's more, \ac{MIIR}-R outperforms Sparse-\ac{MIIR}-R in most cases too.
Modeling the relations between any pair of feature fields helps to make more effective use of item side information to improve sequential recommendation performance.

Comparing \ac{MIIR} with Sparse-\ac{MIIR}, we also notice that the improvement by \ac{DFSA} on the three datasets is higher than that on the discarded versions of the datasets.
A possible reason is that \ac{DFSA} encodes a lot of noisy relations when the missing rate increases.

\subsection{Imputation performance}
\label{RQ4}

To answer RQ4, we compare LRMM and \ac{MIIR} based on their imputation results for the discarded side information.
For different types of feature fields, we consider different metrics:
\begin{enumerate*}[label=(\roman*)]
\item for the category field, we calculate the precision, recall and F1 score for evaluation; 
\item for the brand field, we calculate the accuracy for evaluation; and
\item for the title and description fields, we calculate the mean square error (averaged by the length of title/description vector) for evaluation.
\end{enumerate*}

In Table~\ref{table_5_10}, we list the evaluation results for comparison.
We can observe that \ac{MIIR} achieves better imputation performance than LRMM for the category (in terms of precision), title, and description fields.
But LRMM outperforms \ac{MIIR} for the category (in terms of recall) and brand fields.
Both LRMM and \ac{MIIR} infer discarded side information, so they are able to alleviate the missing side information problem.
Compared with LRMM, \ac{MIIR}  exploits more information from the sequence to impute the missing side information.
However, \ac{MIIR} may also impute some inaccurate results due to over-dependence on the given context.

\subsection{Case study}

\begin{figure}[htbp]
    \centering
    \includegraphics[width=\linewidth]{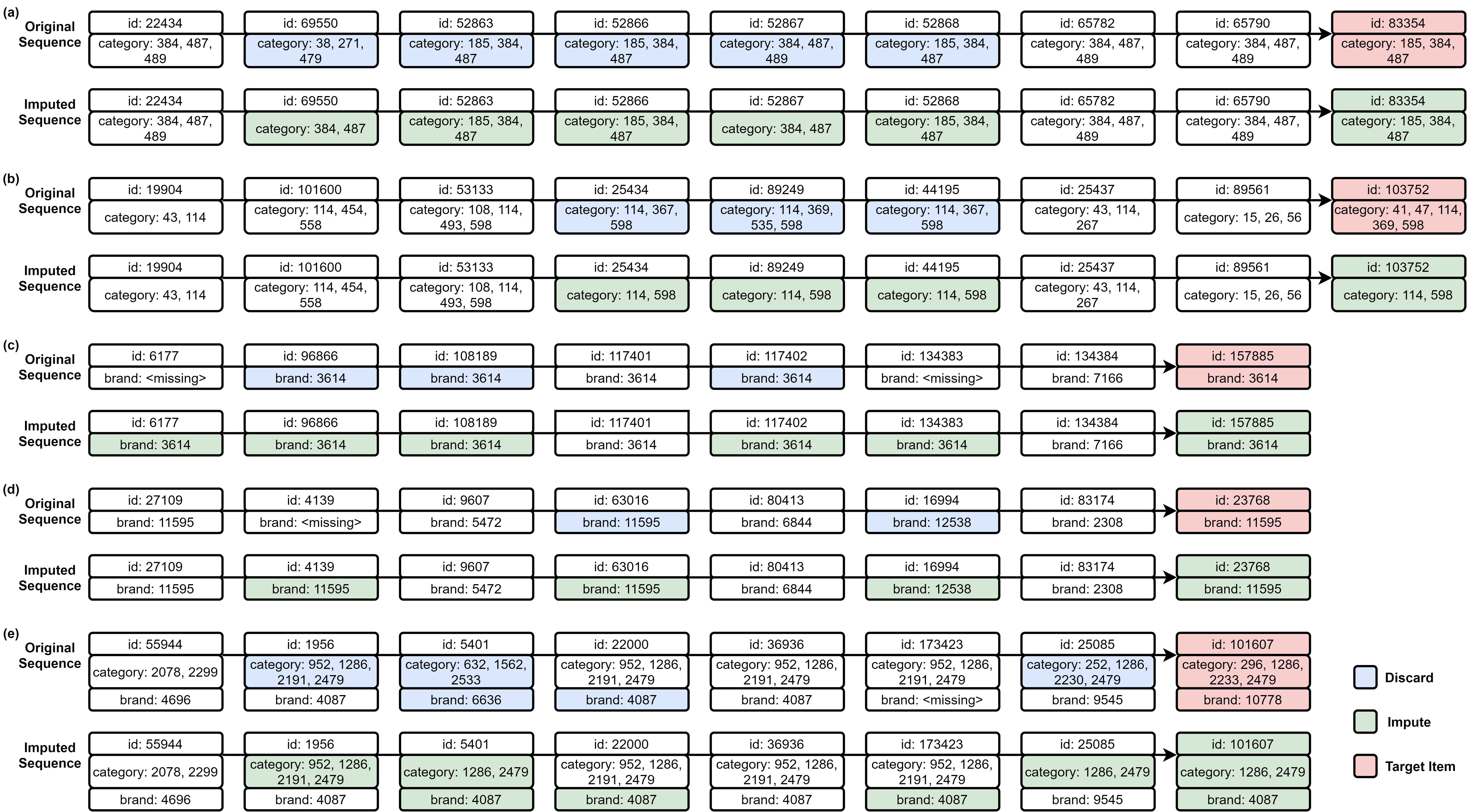}
    \caption{(a) and (b) are two sequences with their imputed categories from the ``Beauty D'' dataset, (c) and (d) are two sequences with their imputed brands from the ``Toys and Games D'' dataset, (e) is the sequence with its imputed categories and brands from the ``Sports and Outdoors D'' dataset.}
    \label{figure_5_1}
\end{figure}

\begin{figure}[t]
    \centering    
    \subfloat[A sequence from the ``Beauty D'' dataset.]{\includegraphics[width=\linewidth]{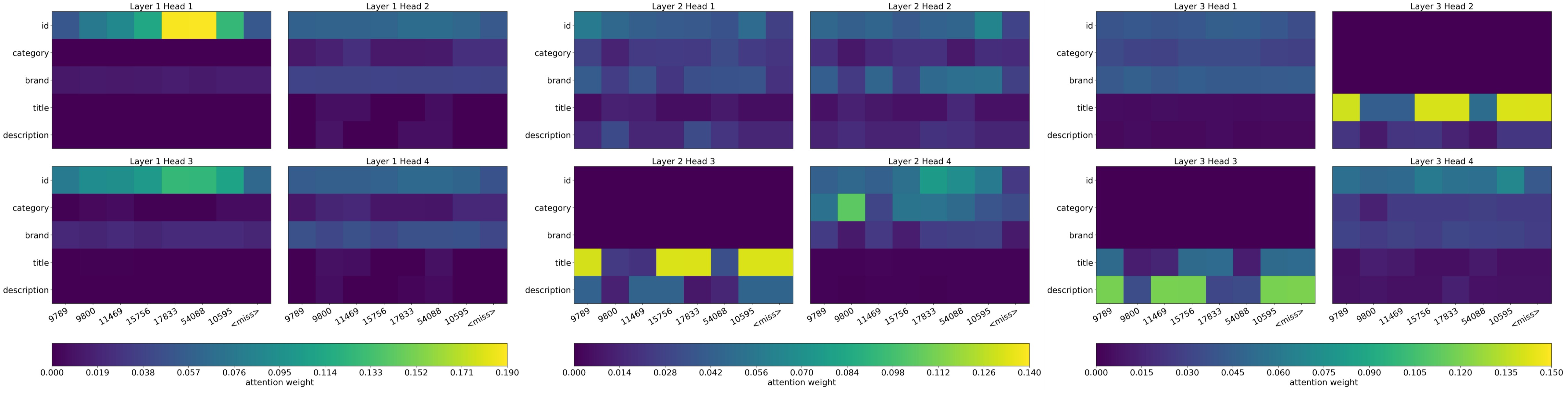}}
    \hfil
    \subfloat[A sequence from the ``Sports and Outdoors D'' dataset.]{\includegraphics[width=\linewidth]{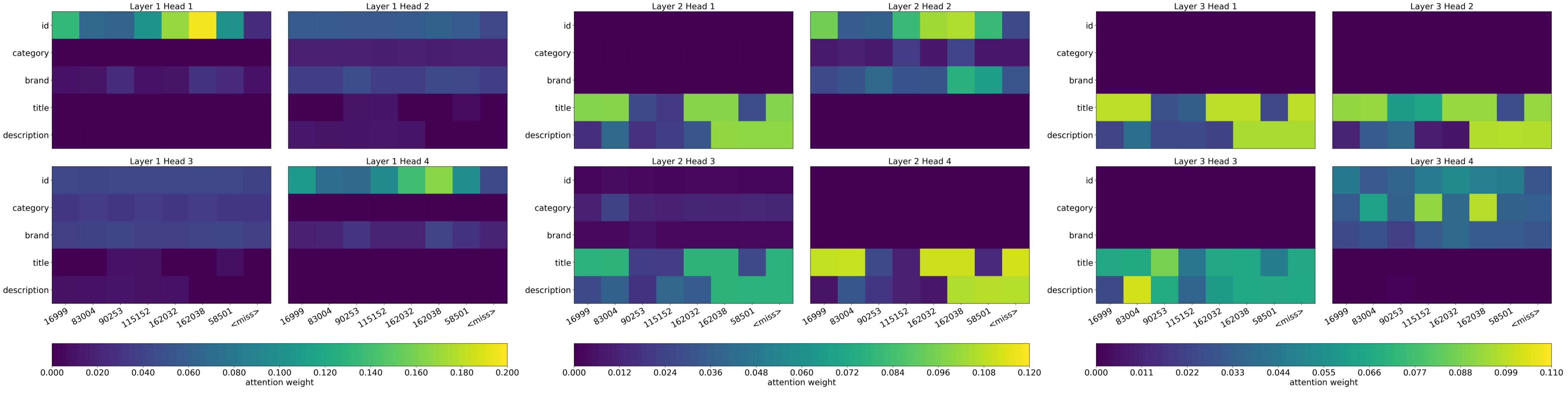}}
    \hfil
    \subfloat[A sequence from the ``Toys and Games D'' dataset.]{\includegraphics[width=\linewidth]{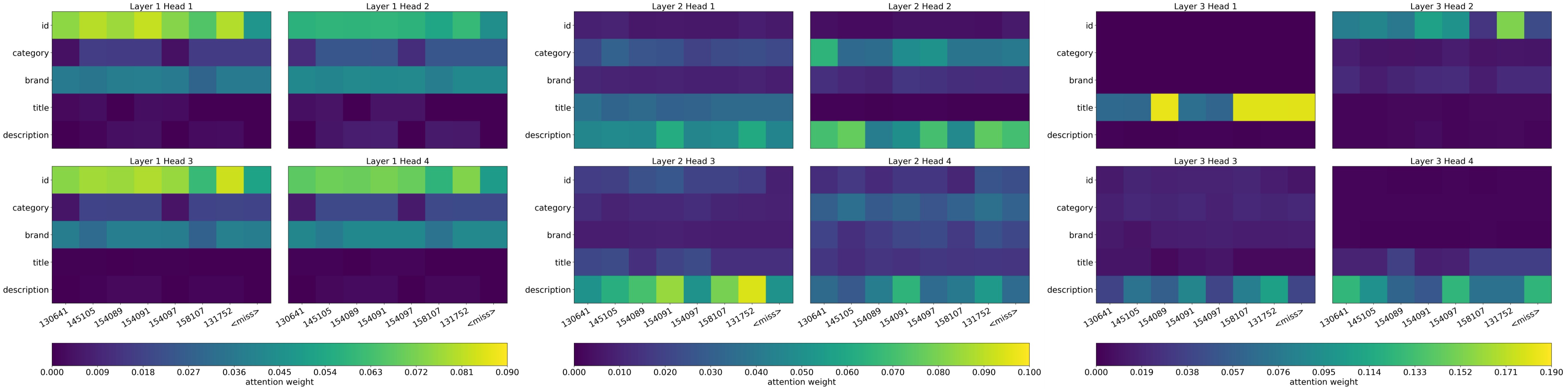}}
    \caption{Visualization for the attention weights from the missing item ID field to all feature fields of all heads and layers in \ac{MIIR} on three sequences from different datasets.}
    \label{figure_5_2}
\end{figure}


In Fig.~\ref{figure_5_1} we list some sequences with their imputed results.
We observe that \ac{MIIR} generates different feature values for missing feature fields according to different contexts (i.e., items and sequences), which is better than using fixed predefined values.
Moreover, \ac{MIIR} is able to infer the ground-truth missing values, including the side information of the next item, to give the model with a more accurate guidance for recommendation.
For example, \ac{MIIR} imputes a part of the discarded categories in sequence (b) and the discarded brands in sequence (d).
We can also observe that the side information of items in the same sequence may be related, which is why \ac{MIIR} can infer the ground-truth missing values in light of the given context.
However, \ac{MIIR} tends to be over-dependent on the information from the sequence, leading it to impute inaccurate results.
For instance, in sequence (e), \ac{MIIR} imputes the wrong categories and brand for item 5401.

Additionally, we visualize the attention weights from the missing item ID (i.e., the next item ID) to all feature fields in the given sequence in \ac{DFSA}, as shown in Fig.~\ref{figure_5_2}.
We reshape the attention weights into a matrix of dimensions $5\times{n}$, where $5$ is the number of the feature field types and $n$ is the sequence length.
First, we see that \ac{MIIR} exploits the information from all feature fields of the given sequence to predict the next item, which emphasizes the necessity to model the relation between any pair of feature fields.
Second, we observe that different layers focus on different types of feature fields, where the first layer mainly attends to ID, and the third layer mainly attends to title and description.
This illustrates that \ac{MIIR} gradually fuses different types of side information into the model by different layers.
Because the information in textual feature fields is more difficult to extract, \ac{MIIR} needs more deeper layers to fuse textual feature fields.
Third, we find that different heads in the same layers have similar attention patterns, which means that there might be some redundant parameters in \ac{MIIR}.

\section{Conclusion}
We have studied the missing side information problem in sequential recommendation. 
We have proposed the \acf{MII} task to unify the missing side information imputation task and the sequential recommendation task.
We have presented a novel sequential recommendation model named \acf{MIIR} to simultaneously impute missing feature values and predict the next item for a given sequence of items.
We have proposed a \acf{DFSA} mechanism to model different relations in the item sequence and to fuse side information.

Based on experiments and analyses on three datasets with different settings of the missing rates we have found that \ac{MIIR} outperforms state-of-the-art methods for sequential recommendation with side information.
We have verified that \ac{MIIR} can identify useful side information from missing feature fields by training with the \ac{MII} task, and that the \ac{DFSA} mechanism improves the recommendation effectiveness of \ac{MIIR}.

As to broader implications of our work, we offer a new perspective by revealing a correlation between missing side information imputation and the sequential recommendation task. They both concern the prediction of missing information.
The perspective operationalized with \ac{MIIR} can be adopted as a foundational paradigm.
Other prediction tasks related to recommendation, such as rating prediction, user profile prediction, and next basket recommendation can also be formulated as a \ac{MII} task.

Limitations of our work are two-fold.
\begin{enumerate*}[label=(\roman*)]
\item Since \ac{DFSA} treats side information as part of the sequence (e.g., in our case, the actual sequence length is 5x the number of items) and models all possible pairwise relations in an item sequence, it is computationally costly and not easy to scale to long sequences; and
\item we have not optimized the \ac{MII} losses on different types of feature fields in \ac{MIIR} for the recommendation task.
\end{enumerate*}

We aim to further improve \ac{MIIR} in different directions.
We will assess the ability of the linear transformer \cite{wang2020linformer,xiong2021nystromformer} to reduce the computational costs of \ac{DFSA} and design a mechanism to filter out useless relations at an early stage.
We also plan to design a tailored loss for \ac{MIIR} by building on recent loss weighting methods \cite{du2018adapting,xu2019multi}.


\begin{acks}
This research was  supported by
the National Key R\&D Program of China with grant (No.2022YFC330 3004, No.2020YFB1406704),
the Natural Science Foundation of China (62102234, 62272274, 62202271, 61902219, 61972234, 62072279),  
the Key Scientific and Technological Innovation Program of Shandong Province (2019JZZY010129), 
the Tencent WeChat Rhino-Bird Focused Research Program (JR-WXG-2021411), 
the Fundamental Research Funds of Shandong University, 
and the Hybrid Intelligence Center, a 10-year program funded by the Dutch Ministry of Education, Culture and Science through the Netherlands Organisation for Scientific Research, \url{https://hybrid-intelligence-centre.nl}.
All content represents the opinion of the authors, which is not necessarily shared or endorsed by their respective employers and/or sponsors.
\end{acks}

\newpage

\bibliographystyle{ACM-Reference-Format}
\bibliography{references}

\end{document}